\documentclass[aps,prd,twocolumn,a4paper,superscriptaddress,10pt,showpacs,notitlepage,floatfix]{revtex4-1}

\usepackage{graphicx}

\usepackage{amsmath}
\usepackage{mathtools}
\usepackage{MnSymbol}
\usepackage{dsfont}
\usepackage{bbold}
\usepackage{slashed}

\usepackage{marvosym}
\usepackage[normalem]{ulem}

\usepackage{lipsum}
\usepackage{xcolor}
\usepackage{colortbl}

\usepackage[plainpages=false,pdfpagelabels,pageanchor=false]{hyperref}

\usepackage[squaren]{SIunits}

\def \beq  {\begin{equation}}
\def \eeq  {\end{equation}}

\newcommand{\overbar}[1]{\mkern 1.5mu\overline{\mkern-1.5mu#1\mkern-1.5mu}\mkern 1.5mu}
\newcommand{\MSbar}{\overbar{\text{MS}}}

\DeclareMathOperator{\Li}{Li}

\newcommand{\xiplus}{\xi_{>}}
\newcommand{\ximinus}{\xi_{<}}

\begin{document}
\allowdisplaybreaks
\begin{abstract}%
High-order behavior of the perturbative expansion for short-distance observables in QCD is 
intimately related to the contributions of small momenta in the corresponding Feynman diagrams
and this correspondence provides one with a useful tool to investigate 
power-suppressed nonperturbative corrections. We use this technique 
to study the structure of power corrections to parton quasi- and pseudo-GPDs 
which are used in lattice calculations of generalized parton distributions. 
As the main result, we predict the functional dependence of the leading power corrections 
to quasi(pseudo)-GPDs on $x$ variable for nonzero skewedness parameter $\xi$.
The kinematic point $x=\pm\xi$ turns out to be special. We 
find that the nonperturbative corrections to quasi-GPDs at this point 
are suppressed by the first power of the hard scale only. 
These contributions come from soft momenta and have nothing 
to do with the known UV renormalon in the Wilson line.  
We also show that power corrections can be strongly suppressed by the normalization procedure.
\end{abstract}

\title{
Renormalons and power corrections in pseudo- and quasi-GPDs}

\author{Vladimir~M.~\surname{Braun}}
\author{Maria~\surname{Koller}}

\affiliation{Institut f{\"u}r Theoretische Physik, Universit{\"a}t Regensburg,\\
Universit{\"a}tsstra{\ss}e 31, 93040 Regensburg, Germany}

\author{Jakob~\surname{Schoenleber}}

\affiliation{Institut f{\"u}r Theoretische Physik, Universit{\"a}t Regensburg,\\
Universit{\"a}tsstra{\ss}e 31, 93040 Regensburg, Germany}

\affiliation{RIKEN BNL Research Center, Brookhaven National Laboratory, Upton, NY 11973, USA} 

\maketitle
\section{Introduction}\label{sect_introduction1}%
Generalized parton distributions (GPDs)~\cite{Mueller:1998fv,Ji:1996ek,Diehl:2003ny,Belitsky:2005qn} have emerged
as an important research object which incorporates the information on the transverse distance 
separation of quark and gluons in the hadron, in dependence on their momentum fractions. 
Their determination is a major  part of ambitious research program at JLAB~\cite{Dudek:2012vr}
and the planned Electron Ion Collider (EIC)~\cite{Accardi:2012qut}
aiming to understand the  proton structure in three dimensions, sometimes referred to as 
``nucleon tomography''. Advances in detector technologies and a very high luminosity of 
these machines  will allow one to study hard exclusive and semi-inclusive reactions with 
identified particles in the final state with unprecedented precision and are expected to make such studies fully 
quantitative.
The Deeply-virtual Compton scattering (DVCS) \cite{Ji:1996nm,Radyushkin:1996nd} is universally accepted as the 
theoretically cleanest reaction that would have the highest potential impact for the transverse distance imaging.
It will be aided by the input from time-like DVCS \cite{Berger:2001xd}, hard exclusive meson production \cite{Brodsky:1994kf} and more sophisticated 
$2\to 3$ exclusive processes \cite{ElBeiyad:2010pji,Boussarie:2016qop,Pedrak:2017cpp,Pedrak:2020mfm,Grocholski:2021man,Grocholski:2022rqj,Duplancic:2018bum,Duplancic:2022ffo,Qiu:2022bpq,Qiu:2022pla,Duplancic:2023kwe}.
The main challenge of these studies is that GPDs are functions of three variables (not counting the scale dependence). 
Their extraction requires massive amount of data and very high precision for both experimental and theory inputs.  

By this reason, any additional information on GPDs from first-principles lattice calculations 
would be extremely important and used in global fits in combination with 
the experimental data, see e.g.~\cite{Riberdy:2023awf} for an exploratory study.
Lattice calculations of the lowest moments of GPDs defined through matrix elements of local composite operators 
have been performed for a long time already and are gaining maturity~\cite{Bali:2018zgl,Alexandrou:2019ali,Alexandrou:2022dtc}.
These calculations are mostly restricted at present to the second moments that are related to the gravitational form factors of the proton and are receiving a lot of attention, see \cite{Burkert:2023wzr} for a recent review.
An alternative approach, originally suggested in \cite{Ji:2013dva}, is to calculate on the lattice 
nucleon (hadron) matrix elements of gauge-invariant nonlocal operators, usually referred 
to as ``quasi'' or ``pseudo'' distributions, qGPDs and pGPDs, respectively. 
For large hadron momenta they can be related to GPDs in 
the framework of collinear factorization in continuum, see~\cite{Radyushkin:2019mye,Ji:2020ect}
for a review and further references.
The specific application of this approach to GPDs was worked out 
in \cite{Ji:2015qla,Xiong:2015nua,Liu:2019urm,Radyushkin:2019owq} and the first
proof-of-the-principle lattice calculations of qGPDs have been performed recently 
\cite{Lin:2020rxa,Alexandrou:2020zbe,Lin:2021brq,Scapellato:2021uke,Bhattacharya:2022aob,Bhattacharya:2023ays}.
Once the main methodical issues for these new methods are established and the lattice simulations 
start moving from exploratory stage towards precision calculations, the 
question of possible size and the structure of the higher-twist (power suppressed) 
corrections to qGPDs and pGPDs has to be addressed.

On the one hand, there exist ``kinematic'' power corrections to p(q)GPDs that are conceptually similar but more complicated
as compared to the well-known Wandzura-Wilczek \cite{Wandzura:1977qf} contributions to the structure functions in polarized
deep-inelastic scattering. Their role is in particular to ensure that GPD extractions from qPDFs and pPDFs  
do not depend on the chosen quark and antiquark positions, the problem noticed in Ref.~\cite{Bhattacharya:2022aob}.  
Kinematic twist-three contributions to  qPDFs and pPDFs  have been scrutinized recently~\cite{Braun:2023alc}
(see also ~\cite{Belitsky:2000vx}) and kinematic twist-four contributions for the {\it moments} of pPDFs and qPDFs
can be  inferred from~\cite{Braun:2011dg}. 

On the other hand, there exist ``genuine'' higher-twist contributions that are 
due to quark and gluon correlations in the hadron and involve new nontrivial nonperturbative input.  
The purpose of this work is study the structure and possible kinematic enhancements of such genuine nonperturbative 
contributions using the concept of renormalons \cite{Beneke:1998ui,Beneke:2000kc}.
This technique exploits the fact that operators of different twist mix with each 
other under renormalization, due to the violation of QCD scale invariance through the running of the coupling constant.
This mixing is explicit in cutoff schemes, whereas in dimensional regularization, it manifests itself in factorial growth 
of the coefficients in the perturbative series at high orders. 
Independence of a physical observable on the factorization scale implies intricate 
cancellations between different twists --- the cancellation of renormalon ambiguities. In turn, the existence of 
these ambiguities in the leading-twist expressions can be used to estimate
the size of power-suppressed corrections. Conceptually, it is similar to the conventional estimation of the accuracy of 
fixed-order perturbative results by the logarithmic scale dependence. 

Renormalons and power corrections to meson pseudo-distribution amplitudes, pseudo- and quasi-PDFs have been considered already 
in Refs.~\cite{Braun:2004bu,Braun:2018brg,Liu:2020rqi}. The present case is more complicated because of the
off-forward kinematics. In particular the point $x=\xi$ on the boundary of the DGLAP and ERBL regions 
requires special treatment and is not directly accessible by the twist expansion. In order to avoid 
proliferation of Lorentz and Dirac structures in this work we consider qGPDs/pGPDs for the spin-zero targets only,
however, the conclusions are expected to be valid for the spin-1/2 targets (nucleon) as well.

The presentation is organized as follows.
The forthcoming Sec.~\ref{sec:definitions} and Sec.~\ref{sec:Borel} are introductory, we collect there 
the necessary definitions and explain our notation. In Sec.~\ref{sec:analytics} we present our results for the Borel transform of the relevant matrix element and
the leading renormalon ambiguities in different representations. The corresponding numerical study employing a simple GPD model 
is carried out in Sec.~\ref{sec:numerics}. Renormalon singularities for qGPDs at the kinematic point 
$x=\xi$  require special treatment and are discussed in Sec.~\ref{sec:x=xi}. 
The final Sec.~\ref{sec:summary} is reserved for a summary and conclusions.  

%
\section{pseudo-GPDs and quasi-GPDs}
\label{sec:definitions}
%

The GPD of a spin-zero particle is defined as an off-forward matrix element of the renormalized leading-twist light-ray operator
\begin{align}
\langle p^\prime|\bar q(\tfrac{z}{2} n)\slashed{n} q(-\tfrac{z}{2} n)|p\rangle
&=
2(Pn) I(\tau,\xi,\Delta^2;\mu)\,, 
\notag\\
I(\tau,\xi,\Delta^2;\mu) &= \int_{-1}^1\!dx\, e^{i\tau x} H(x,\xi,\Delta^2;\mu)\,, 
\label{GPDdef}
\end{align}
where $z\in \mathbb{R}$, $\mu$ is the renormalization scale and $n^\mu$ is an auxiliary light-like 
vector, $n^2=0$. The Wilson line between the quarks is implied.
The relevant kinematic variables are 
\begin{align}
 P= \frac12 (p+p')\,,
&&
\Delta = (p'-p)\,, && \tau = z (Pn) 
\label{kinematics}
\end{align}
and the ``skewedness'' parameter $\xi$ is defined as
\begin{align}
\xi &= \frac{(np)-(np')}{(np)+(np')} = -\frac{(n\Delta)}{2(nP)}\,.
\end{align}
We will assume that the normalization is chosen such that  
$I(\tau=0,\xi=0,\Delta^2=0)=1$ (number of quarks), 
and will not show the dependence on the momentum transfer $\Delta^2$ and the renormalization scale 
$\mu$ in what follows.

GPDs can be determined, at least in principle, from the lattice calculations
of Euclidean correlation functions obtained from (\ref{GPDdef}) by a replacement of the 
light-like vector $n^\mu$ by a space-like vector $v^\mu$,  under the condition that
\begin{align}
 z^2 |v^2| \Lambda_{\rm QCD}^2 \ll 1 \,,\qquad z (vP) = \mathcal{O}(1)\,, 
\end{align} 
which requires that the hadron momentum must be large \cite{Braun:2007wv,Ji:2013dva}.
For definiteness, we assume that $(vP)>0$.  
For scalar targets there exist three Lorentz structures which can be chosen, e.g., as 
\begin{align}
 \langle p^\prime|\bar q(\tfrac{z}{2} v)\gamma^\mu q(-\tfrac{z}{2} v)|p\rangle
& = 2 (vP) \frac{v^\mu}{v^2} \, \mathcal{I}_\parallel(\widetilde\tau,\widetilde\xi,z^2)
\notag\\&\quad
+ 2 \left(P^\mu-(vP) \frac{v^\mu}{v^2}\right)
\mathcal{I}_\perp(\widetilde\tau,\widetilde\xi,z^2)
\notag\\&\quad
+ \Delta_\perp^\mu \mathcal{J}(\widetilde\tau,\widetilde\xi,z^2)
\label{position}
\end{align}
where 
\begin{align}
\widetilde \tau = z (vP)\,, \qquad \widetilde \xi = \frac{(vp) - (vp')}{(vp) + (vp')},
\end{align}
and 
\begin{align}
 \Delta^\mu_\perp = \Delta^\mu + 2\widetilde\xi P^\mu\,, \quad (v\Delta_\perp) = 0\,.
\end{align}
The difference between $\tau$ and  $\widetilde\tau$, and also between $\xi$ and  $\widetilde\xi$
is power suppressed and can be neglected for our purposes.

The three invariant functions in Eq.~\eqref{position} can be separated by
applying  suitable projection operators. Let
\begin{align}
\gamma^\mu_\parallel & =  \frac{v^\mu \slashed{v}}{v^2}
\,, \qquad \gamma^\mu_\perp =   \gamma^\mu - \frac{v_\mu \slashed{v}}{v^2}.
\end{align}
Then obviously
\begin{align}
 \langle p^\prime|\bar q(\tfrac{z}{2} v)\gamma^\mu_\parallel q(-\tfrac{z}{2} v)|p\rangle
&= 2 (vP) \frac{v^\mu}{v^2} \mathcal{I}_\parallel
\notag\\
 \langle p^\prime|\bar q(\tfrac{z}{2} v)\gamma^\mu_\perp q(-\tfrac{z}{2} v)|p\rangle
&= 2  \left(P^\mu-(vP) \frac{v^\mu}{v^2}\right) \mathcal{I}_\perp
+ \Delta_\perp^\mu \mathcal{J}\,.
\label{long-trans}
\end{align}
We will present the results for the both ``longitudinal'' and ``transverse'' invariant functions, 
$\mathcal{I}_\parallel(\widetilde\tau,\widetilde\xi,z^2)$ and $\mathcal{I}_\perp(\widetilde\tau,\widetilde\xi,z^2)$.
The invariant function $\mathcal{J}$  can be expressed in terms of certain integrals involving 
the leading twist GPD (the so-called Wandzura-Wilczek contribution) and the contributions of
twist-three quark-antiquark-gluon light-cone correlation 
functions~\cite{Belitsky:2000vx,Kivel:2000rb,Braun:2023alc}. 
The renormalon contributions to $\mathcal{J}$ correspond to factorial divergences in the 
coefficient functions of quark-antiquark-gluon contributions and are beyond our accuracy. 
Thus for present purposes $\mathcal{J}$ can be omitted.
 
The off-light-cone matrix elements (\ref{position}) can be related to light-cone distributions (GPDs)
using collinear factorization, 
\begin{align}
 \mathcal{I}(\tau,\xi;z^2) &= \int_0^1\!du\, T_{\mathcal I}(u,\tau,\xi,z^2,\mu_F^2)I(u \tau,\xi,\mu_F^2)\,,  
\label{factor-I}
\end{align}
where the coefficient function $T_{\mathcal I}$ can be calculated in perturbation
theory; $\mu_F$ is the factorization scale. At tree level $T_{\mathcal I} = \delta(1-u)$ so that $\mathcal{I}$ 
coincides identically with the position space GPD, 
\begin{align}
 \mathcal{I}(\tau,\xi,z^2) = I(\tau,\xi,\mu^2 \sim 1/z^2|v^2|)
\end{align}
with $\tau = \widetilde\tau$, $\xi = \widetilde\xi$ (up to power corrections).   

Information on the GPDs can be harvested from the measured matrix element (\ref{position}) in several ways. 
Two common approaches are to consider {\it quasi-} and -{\it pseudo} generalized parton distributions 
(qGPDs and pGPDs) defined as 
\begin{align}
\mathcal Q(x,\widetilde \xi; (vP)) &=
(vP) 
\int_{-\infty}^{\infty} \frac{dz}{2\pi} e^{- ixz (vP)} \mathcal I(z (vP), \widetilde \xi; z^2),
\notag\\
\mathcal P(x,\widetilde \xi; z^2) &= 
\int_{-\infty}^{\infty} \frac{d\tau}{2\pi} e^{- i\tau x} \mathcal I(\tau, \widetilde \xi; z^2),
\label{def:qpGPD}
\end{align}
respectively. qPDFs and pPDFs can be related to the GPD by similar factorization theorems that
follow directly from factorization in position space (\ref{factor-I}) and do not need 
additional argumentation.
    
Note that position-space matrix elements (\ref{position}) and hence 
also pGPDs and qPDFs must be renormalized, we do not show the dependence 
on the renormalization scale for brevity. 
The usual logarithmic UV divergences related to multiplicative quark field renormalization
(in axial gauge) are known to three-loop accuracy~\cite{Braun:2020ymy}. In addition, 
in renormalization schemes with an explicit regularization scale, the off-light cone Wilson 
line in Eq.~(\ref{position}) suffers from an additional linear UV divergence that has to be removed.
To this end one can consider a suitable ratio of matrix elements involving the  
Wilson line of the same length, e.g. by normalizing to the same matrix element with $(vP)=0$ and zero skewedness
\cite{Radyushkin:2019owq}. In this way also the logarithmic UV divergences are removed.
We denote the corresponding normalized matrix element by
\begin{align}
\widehat I(\tau, \widetilde \xi; z^2) = \frac{\mathcal I(\tau,\widetilde \xi; z^2)}{\mathcal{I}(0,0;z^2)}.
\label{I-normalized}
\end{align}
This ``ratio method'' is, however, only feasible for ``transverse'' 
p(q)GPDs since $\mathcal{I}_\parallel(0,0;z^2)$ is related to the 
corresponding matrix element by a vanishing prefactor and can only be 
determined via a limiting procedure $(vP)\to 0$, which is impractical.
For completeness we will, nevertheless, present the results for the both 
cases.

The normalized pGPD and qGPD are defined accordingly as
\begin{align}
\widehat Q(x,\widetilde \xi; (vP)) &=
(vP) 
\int_{-\infty}^{\infty} \frac{dz}{2\pi} e^{- ixz (vP)} \widehat I(z (vP), \widetilde \xi; z^2),
\notag\\
\widehat P(x,\widetilde \xi; z^2) &= 
\int_{-\infty}^{\infty} \frac{d\tau}{2\pi} e^{- i\tau x} \widehat I(\tau, \widetilde \xi; z^2)\,.
\label{ratiomethod}
\end{align}
It should be noted that in the existing lattice calculations 
of quasidistributions the ``ratio method'' is not used and the uncertainty due to 
the linear UV divergence of the off-lightcone Wilson line is eliminated using 
different techniques, see \cite{Zhang:2023bxs} for the most recent development.     
The question that we want to address in this work is whether the redefinition in Eq.~\eqref{ratiomethod}
can simultaneously suppress the power corrections coming from the IR region.

%
\section{Borel transform and renormalons}
\label{sec:Borel}
%

The coefficient function $T_{\mathcal I}$ in Eq.~\eqref{factor-I} and the corresponding coefficient functions
for qGPDs and pGPDs in the $\MSbar$ scheme have the perturbative expansion 
\begin{align}
 T = \delta(1-u) + \sum_{k=0}^\infty t_k a_s^{k+1}\,,\qquad a_s = \frac{\alpha_s(\mu)}{4\pi}\,, 
\end{align}
with factorially growing coefficients $t_k\sim k!$.

The standard way to handle such a series is to define the Borel transform
\begin{align}
  B[T] (w) = \sum_{k=0}^\infty \frac{t_k}{k!}\left(\frac{w}{\beta_0}\right)^k 
\end{align}
where powers of $\beta_0 = 11/3 N_c -2/3 n_f$ are inserted for convenience. 
The Borel transform can be viewed as a generating function for the fixed-order coefficients
\begin{align}
 t_k = \beta_0^k \left(\frac{d}{dw}\right)^k  B[T] (w)\big|_{w=0}\,,
\end{align}
and the sum of the series can formally be obtained as the integral 
\begin{align}
\label{BorelIntegral}
T = \delta(1-u)  + \frac{1}{\beta_0} \int_0^\infty \!dw\, e^{-w/(\beta_0 a_s)} B[T](w)\,. 
\end{align}
Note that for one-loop running coupling $ e^{-w/(\beta_0 a_s)} = (\Lambda^2/\mu^2)^{w}$
where $\Lambda \equiv \Lambda_{\rm QCD}^{\MSbar}$. 
As it stands, the integral is usually not defined because the Borel transform has singularities 
on the integration path. 
One can adopt a definition of the integral deforming the contour above or below the real
axis, or as the principle value. These definitions are arbitrary, and their difference, which is exponentially small in the 
coupling, must be viewed as an intrinsic uncertainty of perturbation theory that has to be removed by adding 
power-suppressed nonperturbative corrections. In this sense, the factorial divergences are not 
a problem by themselves, but rather a messenger of a problem: existence of nonperturbative 
contributions to correlation functions that are missed by perturbation theory. Studying 
the factorial divergences we can, therefore, reveal useful information on the missing nonperturbative effects.

Naturally, a full all-order calculation cannot be performed. 
It can be argued~\cite{Beneke:1998ui,Beneke:2000kc} that the most important (closest to the origin) singularities 
of the Borel transform of QCD correlation functions can be traced by  
computing the diagrams with the insertion of multiple fermion loops in the one-loop diagram
and replacing  $- \frac23 n_f \mapsto   \beta_0 = \frac{11}{3} N_c - \frac23 n_f$.
Such singularities are intuitively related to the running-coupling effects and are usually 
referred to as ``renormalons'' \cite{tHooft:1977xjm}.
 
For the case at hand, one expects the structure of singularities of the Borel transform of the 
coefficient functions as illustrated in Fig.~\ref{fig:borelplane}: 
\begin{figure}[h]
\centering
 \includegraphics[width=0.45\textwidth]{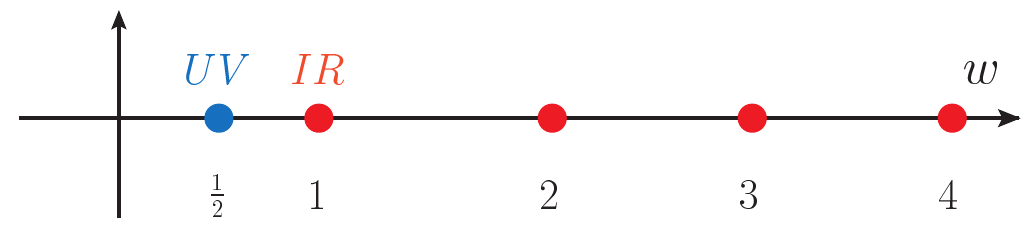}
\caption{Typical singularity structure of the Borel transform.}
\label{fig:borelplane}
\end{figure}
There is an UV renormalon singularity at $w=1/2$ related to the linear UV divergence of the Wilson line, 
and a series of IR renormalons at positive integer $w=1,2,\ldots$
that must be matched by power-suppressed (higher-twist) nonperturbative corrections. 
In the single bubble chain approximation all singularities are usually simple poles.
We will justify this picture for pGPDs by explicit calculation, whereas for qGPDs at 
the special kinematic point $x=\xi$ the situation proves to be somewhat more complicated.

\begin{figure}[h]
\centering
 \includegraphics[width=0.42\textwidth]{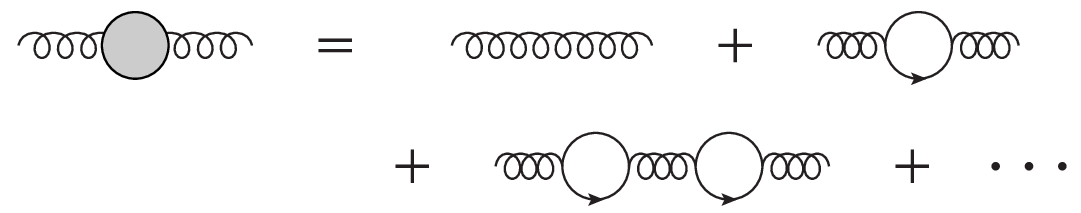}
\caption{Bubble-chain contribution to the gluon propagator.}
\label{fig:bubblechain}
\end{figure}

In sufficiently simple cases a Borel transform can be calculated directly. 
Evaluation of diagrams with multiple fermion bubble insertions in Landau gauge
corresponds to the evaluation of the lowest-order diagram with the effective 
gluon propagator, Fig.~\ref{fig:bubblechain},
\begin{align}
D_{\mu\nu}^{AB} &= \frac{i\delta^{AB}}{-k^2-i\epsilon}\left(g_{\mu\nu}- \frac{k_\mu k_\nu}{k^2}\right)
\frac{1}{1+\Pi(k^2)}\,, 
\label{bubblesum}
\end{align}
where in the  $\MSbar$ scheme
\begin{align}
 \Pi(k^2) &= \beta_0 a_s \ln\left(-\frac{k^2}{\mu^2} e^{-5/3}\right)\,.
\end{align}

For the diagrams with only one bubble chain the Borel transformation effectively applies to the expansion
in $\alpha_s$ of the propagator in Eq.~\eqref{bubblesum} rather than to the diagrams as a whole.
The effective Borel-transformed propagator is \cite{Beneke:1992ch}
\begin{align}
B[a_s D^{AB}(k)](w) = i\delta^{AB} \left(e^{5/3} \mu^2\right)^{w} 
\frac{g_{\mu\nu}- \frac{k_\mu k_\nu}{k^2}}{(-k^2-i\epsilon)^{1+w}}\,.
\label{gluonprop}
\end{align}
It is easy to see that using this expression is equivalent to replacing the (one-loop) running coupling 
constant in the lowest-order diagram by an effective coupling 
\begin{align}
  \beta_0 a_s(-k^2) =  \int_0^\infty \!dw\, e^{\frac53 w} \left(\frac{\Lambda^2}{-k^2}\right)^w .
\end{align}
This replacement leads to the modified form of the gluon propagator 
\begin{align}\label{modgluonprop}
  \frac{1}{-k^2-i\epsilon} ~\mapsto~ \frac{\Lambda^{2w}}{(-k^2-i\epsilon)^{1+w}}\,,
\end{align}
which is familiar from analytic regularization \cite{Speer:1975gj}. Note that for gauge-invariant quantities the 
requirement of using Landau gauge is immaterial, Feynman gauge can be used instead. 

Alternatively, the desired information on the high-order behavior can be obtained 
from the lowest-order diagrams, calculated with a finite gluon mass~\cite{Beneke:1994qe,Ball:1995ni}.
A formal equivalence of these two methods can be established by the (inverse) Mellin transform
\begin{align}
 \frac{1}{k^2-\lambda^2} &= 
\frac{1}{2\pi i} \frac{1}{k^2} 
\int\limits_{-\frac12-i\infty}^{-\frac12+i\infty}
\!\!\!dw \,\Gamma(-w)\Gamma(1+w)\left(-\frac{\lambda^2}{k^2}\right)^w.
\label{gluonmass}
\end{align}
The renormalon singularities correspond in this approach to {\it nonanalytic}
terms in the expansion of the amplitude in powers of the gluon mass, e.g., 
terms  $\sim \lambda^2\ln\lambda^2$ or $\sqrt{\lambda^2}$.
The advantage of this technique is that it allows one to get a 
simple integral representation for the contribution of $k$ bubbles and the 
bubble sum (e.g. with principal value prescription), which is not easy 
starting from the Borel transform. 
We have used both methods for a cross-check of the results.

\begin{figure*}[htb!]
\centering
 \includegraphics[width=0.95\textwidth]{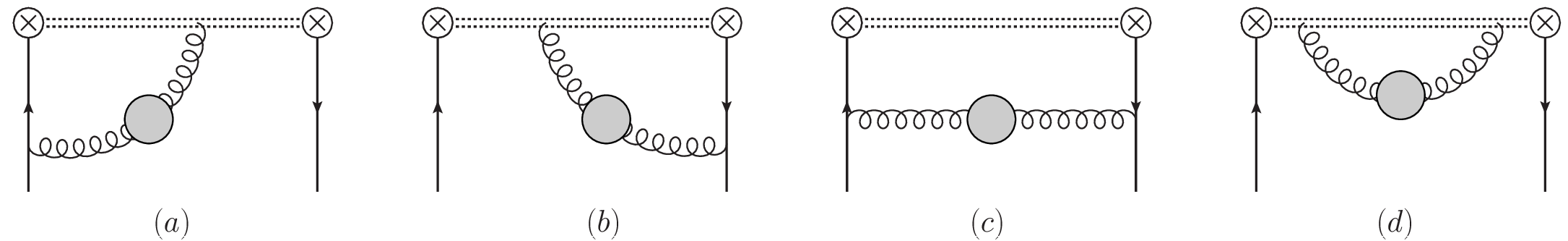}
\caption{Bubble-chain contribution to pGPDs/qGPDs. The Wilson line connecting the quark fields is shown by double dots.}
\label{fig:diagrams}
\end{figure*}
%

%
\section{Renormalon ambiguities}
\label{sec:analytics}
%

In this section we present analytic expressions for the renormalon ambiguities in the matrix element \eqref{position} in 
position space, and the corresponding pGPD/qGPD.
Replacing the gluon propagator in the one-loop diagrams in Fig.~\ref{fig:diagrams} 
by the Borel transform of the bubble chain \eqref{gluonprop} one obtains 
\begin{widetext}
\begin{align}
B[\mathcal{I}(\tau,\xi,z^2)](w) & =
2 C_F e^{5/3w} \left(\frac{-z^2 v^2\mu^2}{4 }+i0\right)^w \frac{\Gamma(-w)}{\Gamma(w+1)}   
\notag\\&\quad
\times 
\biggl\{ \frac{2}{1+w} \int_0^1 d\alpha\,\alpha^{1+w}  {}_2F_1(1,2-w,2+w,\alpha)
\Big[
\cos(\bar\alpha\xi\tau) I(\alpha \tau,\xi;\mu) - I(\tau,\xi;\mu)
\Big] 
\notag\\&\quad
- \frac{1}{1+w} I(\tau,\xi;\mu) - \frac{1}{1-2w} I(\tau,\xi;\mu) 
+ \big[1\mp w] \int_0^1\!d\alpha\,\alpha^w \frac{\sin(\bar\alpha\xi\tau)}{\tau\xi} I(\alpha \tau,\xi;\mu)  
\biggr\}, 
\label{start}
\end{align}
\end{widetext}
where in the last term the ``minus'' sign corresponds to $\mathcal{I}_\parallel$ and the ``plus'' sign to
$\mathcal{I}_\perp$, respectively. Here and below we use the notation  
\begin{align}
\bar\alpha = 1-\alpha\,.
\end{align}
The calculation is straightforward and follows closely Ref.~\cite{Braun:2004bu} so that we omit the details.
Note that the expression in Eq.~\eqref{start} corresponds to the calculation with the chain of {\it renormalized} bubbles,
but the overall subtraction of the logarithmic divergence in the one-loop diagram is not yet done.
The singularity of the Borel transform in Eq.~\eqref{start} at $w=0$ is removed by this subtraction.
(The coefficient of the $1/w$ pole is nothing but the evolution kernel for the GPDs in position space
 representation.)
Importantly, the overall subtraction in minimal subtraction schemes does not 
affect~\cite{Beneke:1994sw,Beneke:1998ui} renormalon singularities located further to the right in the Borel plane, 
which are subject of this study. Thus we do not need to perform this subtraction explicitly.

The closest to the origin singularity at $w = 1/2$ is generated by the 
contribution of large momenta in the self-energy insertions in the Wilson line, diagram Fig.~\ref{fig:diagrams}d,
 and is part of the UV renormalization factor.
This singularity is well-known and has to be removed by a suitable renormalization procedure,
see, e.g., Ref.~\cite{Zhang:2023bxs}.
It will not be considered further in this work. 

The leading IR renormalon singularity is at $w=1$.
An ambiguity in the sum of a fixed-sign factorially-divergent series 
is usually estimated by the imaginary part (divided by $\pi$) of the corresponding Borel integral \eqref{BorelIntegral}. This ambiguity is canceled exactly by the UV renormalon in the matrix elements of the contributing 
twist-four operators, see e.g.~\cite{Braun:2004bu}. 
Following the standard logic~\cite{Beneke:1998ui,Beneke:2000kc} we assume that the ``true'' nonperturbative
correction  that is left after the renormalon cancellation, is of the same order of magnitude. 

\subsection{Position space}
From Eq.~\eqref{start} we obtain
\begin{align}
 \mathcal{I}(\tau, \xi, z^2) &= I(\tau,\xi)
\pm 
\mathcal{N}\,\Big(\Lambda^2z^2|v^2|\Big)
\delta_R {\mathcal{I}}(\tau,\xi)\,,
\notag\\[2mm]
 \widehat{\mathcal{I}}(\tau, \xi, z^2) &= I(\tau,\xi)
\pm 
\mathcal{N}\,\Big(\Lambda^2z^2|v^2|\Big)
\delta_R\widehat{\mathcal{I}}(\tau,\xi)\,, 
\label{Ihat}
\end{align}
with
\begin{align}
 \delta_R {\mathcal{I}}^\parallel(\tau,\xi) &= 
\int_0^1\!d\alpha\, I(\alpha \tau,\xi)\,\Phi(\alpha)\,\cos(\bar\alpha\tau\xi)\,,
\notag\\
 \delta_R \widehat{\mathcal{I}}^\parallel(\tau,\xi) &= 
\int_0^1\!d\alpha\, I(\alpha \tau,\xi)\,[\Phi(\alpha)]_+\,\cos(\bar\alpha\tau\xi)
\notag \\ & =  \delta_R {\mathcal{I}}^\perp(\tau,\xi) - \frac{1}{4} I(\tau,\xi)\,,
\end{align}
and
\begin{align}
 \delta_R {\mathcal{I}}^\perp(\tau,\xi) &=  \delta_R \mathcal{I}^\parallel(\tau,\xi)
+ \int_0^1\!d\alpha\, I(\alpha \tau,\xi)\,\alpha \frac{\sin(\bar\alpha\tau\xi)}{\xi\tau}\,,
\notag\\
\delta_R \widehat{\mathcal{I}}^\perp(\tau,\xi) &= 
 \delta_R \widehat{\mathcal{I}}^\parallel(\tau,\xi)
+ \int_0^1\!d\alpha\, I(\alpha \tau,\xi)
 [\alpha\bar\alpha]_+ \frac{\sin(\bar\alpha\tau\xi)}{\bar\alpha\xi\tau}
\notag\\ &=  \delta_R {\mathcal{I}}^\perp(\tau,\xi) - \frac{5}{12} I(\tau,\xi).
\end{align}
%
where
\begin{align}
 \Phi(\alpha) &= \alpha + \bar\alpha \ln\bar \alpha\,,
&&
\mathcal N =\left[\frac{C_F e^{5/3}}{\beta_0}\right] \simeq 0.76
\end{align}
and the ``plus'' distribution is defined as
\begin{align}
  [f(\alpha)]_+ = f(\alpha) - \delta(\bar\alpha)\int_0^1 d\beta\, f(\beta)\,. 
\end{align}
For zero skewedness, $\xi=0$, these expressions coincide  with Ref.~\cite{Braun:2018brg} apart from a  different overall normalization convention.

\subsection{pseudo-GPDs}
The results for pGPDs follow from the above expressions by the  
Fourier transform in $\tau$ for fixed $z^2$. We write
\begin{align}
 \mathcal{P}(x, \xi, z^2) &= H(x,\xi)
\pm 
\mathcal{N}\,\Big(\Lambda^2z^2|v^2|\Big)
\delta_R {\mathcal{P}}(x,\xi)\,,
\notag\\
 \widehat{\mathcal{P}}(x, \xi, z^2) &= H(x,\xi)
\pm 
\mathcal{N}\,\Big(\Lambda^2z^2|v^2|\Big)
\delta_R \widehat{\mathcal{P}}(x,\xi)\,, 
\label{pGPD-ambiguity}
\end{align} 
and obtain after a short calculation
\begin{widetext}
\begin{align}
2 \delta_R {\mathcal{P}}^\parallel(x,\xi)&=
\theta(x>\xi)
\biggl[
\int_x^1dy \frac{H(y,\xi)}{y-\xi} \Phi\left(\frac{x-\xi}{y-\xi}\right)
+ 
\int_x^1dy \frac{H(y,\xi)}{y+\xi} \Phi\left( \frac{x+\xi}{y+\xi}\right)
\biggr]
\notag\\&+
\theta(-\xi<x<\xi)
\biggl[
 - \int_{-1}^xdy \frac{H(y,\xi)}{y-\xi} \Phi\left( \frac{x-\xi}{y-\xi}\right)
+ 
\int_x^1dy \frac{H(y,\xi)}{y+\xi} \Phi\left( \frac{x+\xi}{y+\xi}\right)
\biggr]
\notag\\&+
\theta(x<-\xi)
\biggl[
 - \int_{-1}^xdy \frac{H(y,\xi)}{y-\xi} \Phi\left(\frac{x-\xi}{y-\xi}\right)
-
\int_{-1}^xdy \frac{H(y,\xi)}{y+\xi} \Phi\left(\frac{x+\xi}{y+\xi}\right) 
\biggr]
\notag\\
2 \delta_R {\mathcal{P}}^\perp(x,\xi)&= 2 \delta_R {\mathcal{P}}^\parallel(x,\xi) 
 + \frac{1}{2\xi} 
\biggl\{ 
 \theta(x>\xi) \int_{x}^1\!dy\, \left[ \left(\frac{x+\xi}{y+\xi}\right)^2 - \left(\frac{x-\xi}{y-\xi}\right)^2\right]  \, H(y,\xi)
\notag\\&\qquad
+\theta(-\xi < x < \xi)\biggl[ \int_{x}^1\!dy\, \left(\frac{x+\xi}{y+\xi}\right)^2\, H(y,\xi)
+ 
\int_{-1}^x\!dy\,\left(\frac{x-\xi}{y-\xi}\right)^2\, H(y,\xi)\biggr]
\notag\\&\qquad
+ \theta(x<-\xi) \int_{-1}^x dy \left[ \left(\frac{x-\xi}{y-\xi}\right)^2 - \left(\frac{x+\xi}{y+\xi}\right)^2\right]  \, H(y,\xi)
\biggr\}
\label{pGPD-renormalons}
\end{align}
\end{widetext}
and for the normalized pGPDs
\begin{align}
 \delta_R \widehat{\mathcal{P}}^\parallel(x,\xi)&=
 \delta_R {\mathcal{P}}^\parallel(x,\xi) - \frac14 H(x,\xi)\,,
\notag\\
 \delta_R \widehat{\mathcal{P}}^\perp(x,\xi)&=
 \delta_R {\mathcal{P}}^\perp(x,\xi) - \frac{5}{12} H(x,\xi)\,.
\label{hat-pGPD}
\end{align}

Note that the (quark) GPD $H(x,\xi)$ is a continuous function of $x$ at $x=\xi$, but its derivatives are, generally, discontinuous.
One obtains
\begin{align}
 \int_x^1dy \frac{H(y,\xi)}{y-\xi} \Phi\left(\frac{x-\xi}{y-\xi}\right)\Big|_{x\to\xiplus} &= \left(2-\frac{\pi^2}{6}\right)H(\xi,\xi)\,,
\notag\\
\int_{-1}^xdy \frac{H(y,\xi)}{y-\xi} \Phi\left( \frac{x-\xi}{y-\xi}\right)\Big|_{x\to\ximinus}  &= - \left(2-\frac{\pi^2}{6}\right)H(\xi,\xi)\,,
\end{align}
where
\begin{align}
 \xiplus = \xi+\epsilon\,, && \ximinus = \xi-\epsilon\,,&& \epsilon \to 0\,,
\end{align}
and it is easy to check that the renormalon contribution to the pGPDs is continuous at $x=\xi$.

The limiting case $\xi=1$ corresponds to the kinematics that is relevant for (pseudo) light-cone 
distribution amplitudes (LCDAs). Choosing $H(x,\xi=1) = \phi_\pi(x) = 3/4 (1-x^2)$ (asymptotic pion LCDA) 
we obtain for the corresponding renormalon ambiguity

\begin{align}
 \delta_R \mathcal{I}_{\pi}^{\parallel}(x)
&=   \frac16 \Big\{
 \bar u [\ln\bar u -\Li_2(\bar u)] + u[\ln u -\Li_2(u)]
\notag\\&\quad  - u\bar u + \frac{\pi^2}{6} 
\Big\}\Big|_{u = \frac12(1+x)},
\end{align}
\noindent
in agreement with Eq.(3.8) in \cite{Braun:2004bu}. 
The sign and the overall normalization of the corresponding nonperturbative correction 
can in this case be adjusted to the known twist-four matrix element, see \cite{Braun:2004bu,Bali:2018spj} for details.\\[2mm]

\subsection{quasi-GPDs}

It proves to be convenient to write the renormalon ambiguity 
for qGPDs in terms of the derivative
\begin{align}
 H'(x,\xi) = \frac{d}{dx} H(x,\xi)\,. 
\end{align}
We obtain
\begin{align}
 \mathcal{Q}(x, \xi, (vP)) &= H(x,\xi)
\pm 
\mathcal{N}\,\left(\frac{\Lambda^2|v^2|}{(vP)^2}\right)
\delta_R {\mathcal{Q}}(x,\xi)\,,
\notag\\
 \widehat{\mathcal{Q}}(x, \xi,(vP)) &= H(x,\xi)
\pm 
\mathcal{N}\,\left(\frac{\Lambda^2|v^2|}{(vP)^2}\right)
\delta_R \widehat{\mathcal{Q}}(x,\xi)\,, 
\label{qGPD-ambiguity}
\end{align} 
with

\begin{figure*}[htb!]
\centering
 \includegraphics[width=0.32\textwidth]{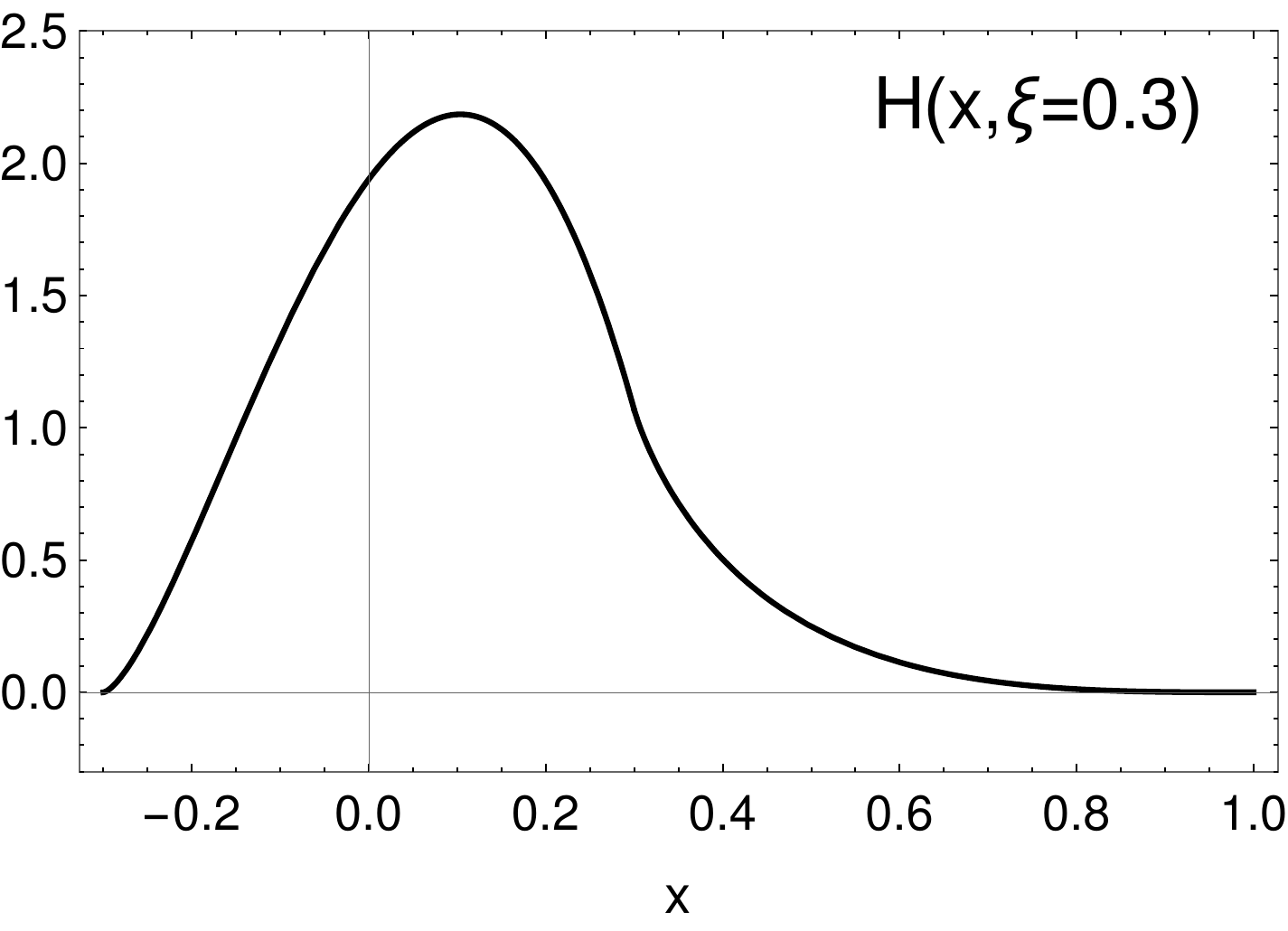}
 \includegraphics[width=0.32\textwidth]{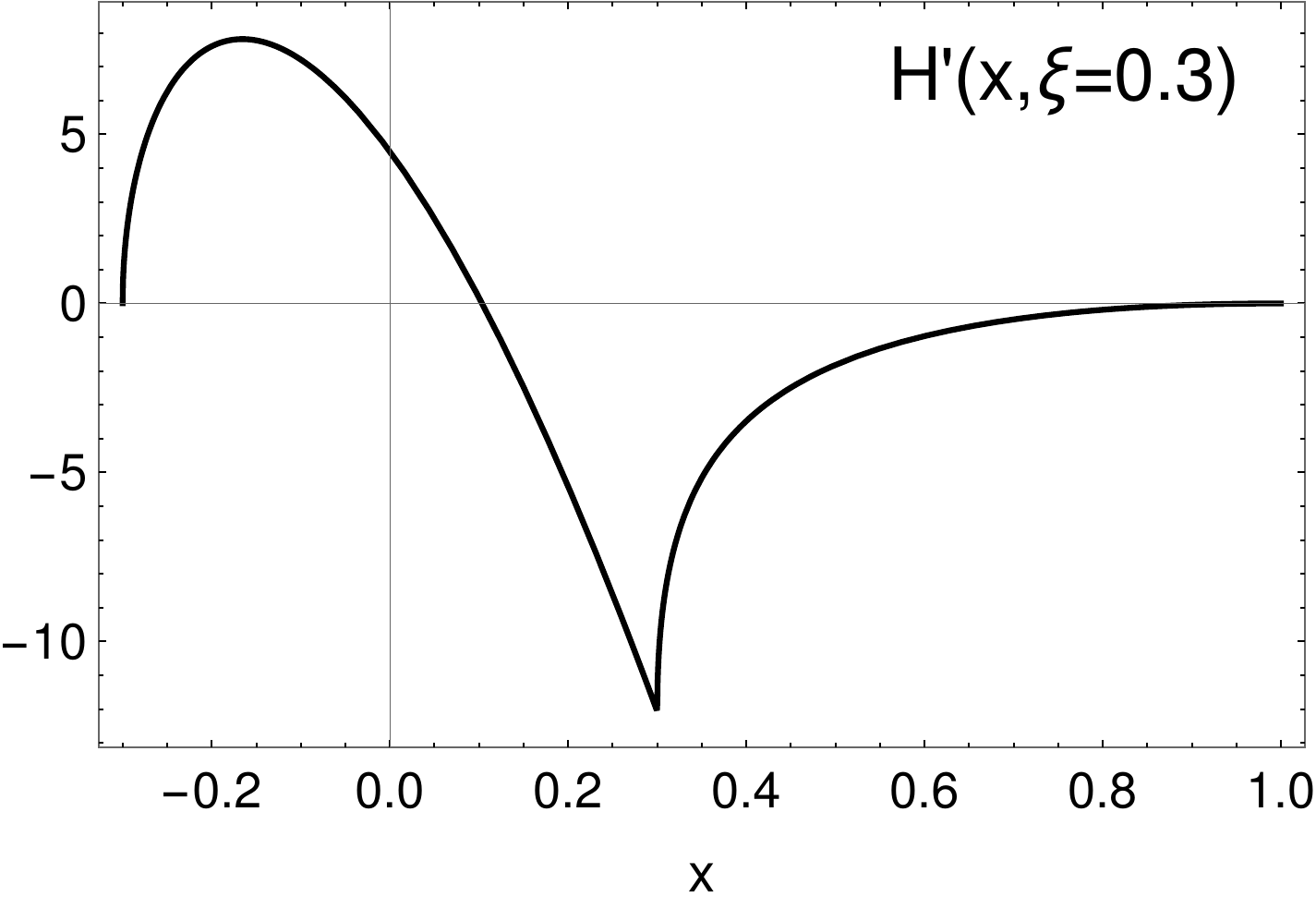}
 \includegraphics[width=0.32\textwidth]{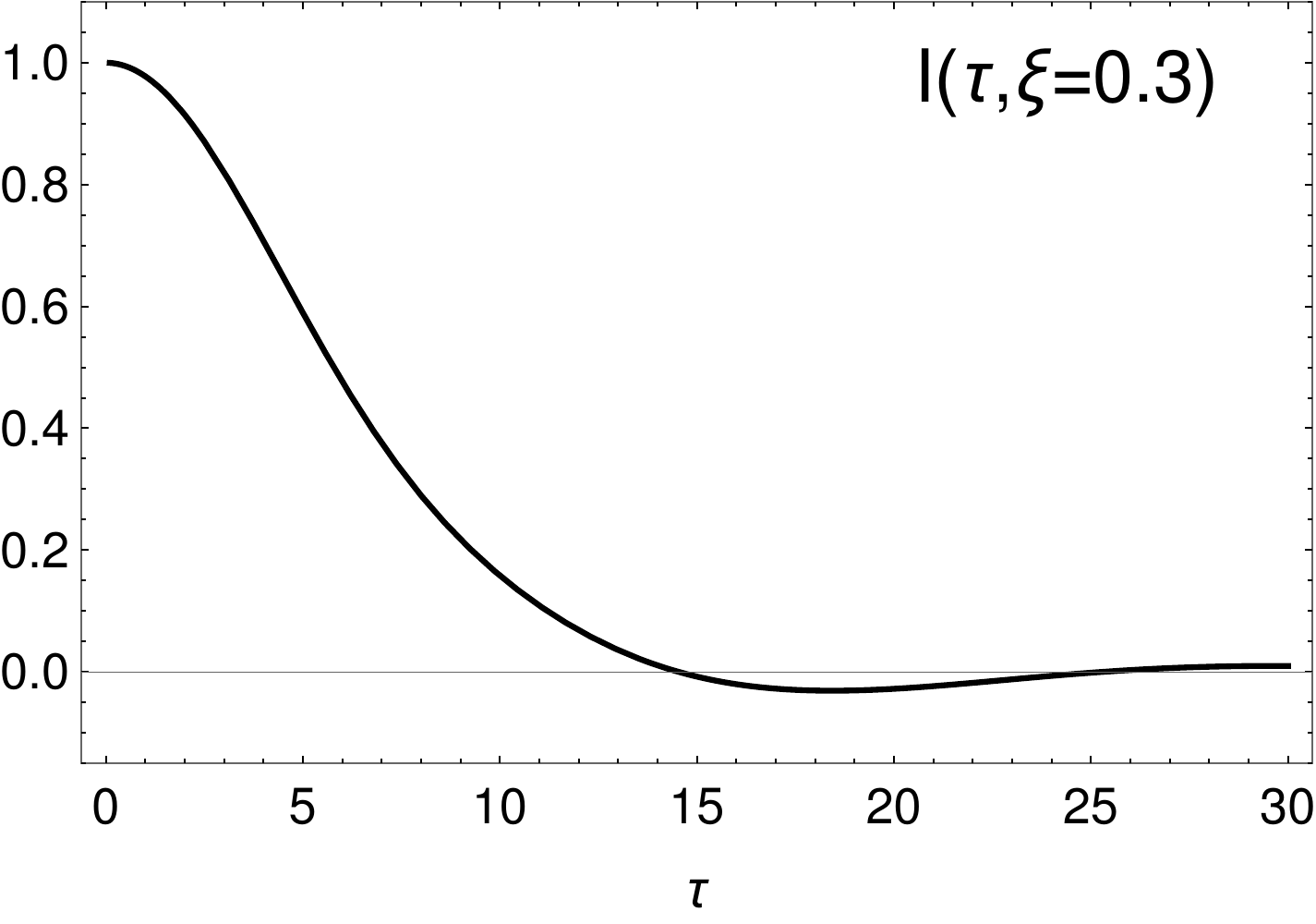}
\caption{GPD model \eqref{GPDmodel} for $\xi=0.3$ used in our calculations: $H(x,\xi)$ and the derivative $H'(x, \xi)$ 
as functions of $x$ (left and middle panels, respectively), 
and the real part of the corresponding  position space GPD $I(\tau,\xi)$ a function of $\tau$ (right panel).}
\label{fig:GPDmodel}
\end{figure*}

\begin{widetext}
\begin{align}
2 \delta_R {\mathcal{Q}}^\parallel(x,\xi)
&= 
\left(1-\frac{\pi^2}{6}\right) \delta(x-\xi) [H'(\xiplus,\xi)- H'(\ximinus,\xi)] 
\notag\\&\quad
+ \theta(x>\xi)\left[ \frac{H'(x,\xi)}{x-\xi} + \frac{1}{(x-\xi)^2} \int_{x}^1\!dy\,\left[\frac{x-\xi}{y-\xi}+\ln\left(1-\frac{x-\xi}{y-\xi}\right)\right] \, H'(y,\xi)\right] 
\notag\\&\quad 
+ \theta(x<\xi)\left[\frac{H'(x,\xi)}{x-\xi}  -  \frac{1}{(x-\xi)^2} \int_{-1}^x\!dy\, \left[\frac{x-\xi}{y-\xi}+\ln\left(1-\frac{x-\xi}{y-\xi}\right)\right] \, H'(y,\xi)\right]  
\notag\\&\quad
+
\left(1-\frac{\pi^2}{6}\right) \delta(x+\xi) [H'(-\ximinus,\xi)- H'(-\xiplus,\xi)] 
\notag\\&\quad
+ \theta(x>-\xi)\left[ \frac{H'(x,\xi)}{x+\xi} + \frac{1}{(x+\xi)^2} \int_{x}^1\!dy\,\left[\frac{x+\xi}{y+\xi}+\ln\left(1-\frac{x+\xi}{y+\xi}\right)\right] \, H'(y,\xi)\right] 
\notag\\&\quad 
+ \theta(x<-\xi)\left[\frac{H'(x,\xi)}{x+\xi}  -  \frac{1}{(x+\xi)^2} \int_{-1}^x\!dy\, \left[\frac{x+\xi}{y+\xi}+\ln\left(1-\frac{x+\xi}{y+\xi}\right)\right] \, H'(y,\xi)\right], 
\notag\\
2\delta_R {\mathcal{Q}}^\perp(x,\xi)
& =2 \delta_R {\mathcal{Q}}^\parallel(x,\xi) 
+
 \frac{2}{2\xi} 
\biggl\{
 \theta(x>\xi)\int_{x}^1\!dy\,\frac{ H'(y,\xi)}{(y-\xi)}
-  \theta(x<\xi) \int_{-1}^x\!dy\, \frac{ H'(y,\xi)}{(y-\xi)}
\notag\\&\hspace*{1.7cm}
-  \theta(x>-\xi)\int_{x}^1\!dy\,\frac{ H'(y,\xi)}{(y+\xi)}
+ \theta(x<- \xi) \int_{-1}^x\!dy\, \frac{ H'(y,\xi)}{(y+\xi)}
\biggr\}.
\label{qGPD-renormalons}
\end{align}
\end{widetext}
Note appearance of the $\delta(x\pm\xi)$ contributions proportional to the discontinuity of the derivative $H'(x,\xi)$ at 
$x=\pm \xi$. It is easy to verify that these contributions arise from large distances in the Fourier integral 
so that a small-$z^2$ expansion in position space is not sufficient.
In this special kinematics, one has to search for renormalon singularities in qGPDs directly,  after the Fourier 
transformation to momentum space. We postpone this calculation and the corresponding discussion to   
Sec.~\ref{sec:x=xi}. 

For the normalized qGPDs we obtain
\begin{align}
\label{deltahatQ}
\delta_R \widehat{\mathcal Q}^{\parallel}(x,\xi) &= \delta_R\mathcal Q^{\parallel}(x,\xi) + \frac{1}{4} H''(x,\xi)\,, 
\notag\\
\delta_R \widehat{\mathcal Q}^{\perp}(x,\xi) &= \delta_R \mathcal Q^{\perp}(x,\xi) + \frac{5}{12}H''(x,\xi),
\end{align}
where the second derivative $H''(x,\xi)$ contains delta function contributions 
at $x\pm\xi$ where $H'(x,\xi)$ is discontinuous:
\begin{align}
 H''(x,\xi) &= \ldots + 
\delta(x-\xi) [H'(\xiplus,\xi)- H'(\ximinus,\xi)] 
\notag\\[1mm] &\quad 
+  \delta(x+\xi) [H'(-\ximinus,\xi)- H'(-\xiplus,\xi)]\,. 
\end{align}

%
\section{Numerical study}
\label{sec:numerics}
%

\subsection{GPD model}

We will use a simple model for the valence quark GPD
based on the standard double-distribution ansatz~\cite{Radyushkin:1997ki,Belitsky:2005qn}
\begin{align}
H(x,\xi) & =  \theta(x>-\xi) \frac{2+\lambda}{4 \xi^3} \left(\frac{x+\xi}{1+\xi}\right)^{\lambda} [\xi^2 -x + \lambda \xi (1-x)]
\notag\\&\quad
- \theta(x>\xi) \frac{2+\lambda}{4\xi^3} \left(\frac{x-\xi}{1-\xi}\right)^{\lambda} [\xi^2 -x - \lambda \xi (1-x)]\,,
\label{GPDmodel}
\end{align}
with $\lambda= 3/2$ corresponding to the valence quark PDF 
$q(x) \sim (1-x)^3 x^{\lambda-2} \equiv (1-x)^3/\sqrt{x}$.
This model, the $x$-derivative $H'(x,\xi)$,
and the real part of the corresponding position space GPD 
$I(\tau,\xi)$ are shown in Fig.~\ref{fig:GPDmodel} for $\xi=0.3$ as an illustration.

Note that the valence quark GPD vanishes for $x<-\xi$. 
For the chosen values of parameters $H(x,\xi) \sim (1-x)^3$ for $x\to 1$ and 
 $H(x,\xi) \sim (x+\xi)^\lambda \equiv (x+\xi)^{3/2} $ for $x\to -\xi$.

\begin{figure*}[htb!]
\centering
 \includegraphics[width=0.32\textwidth]{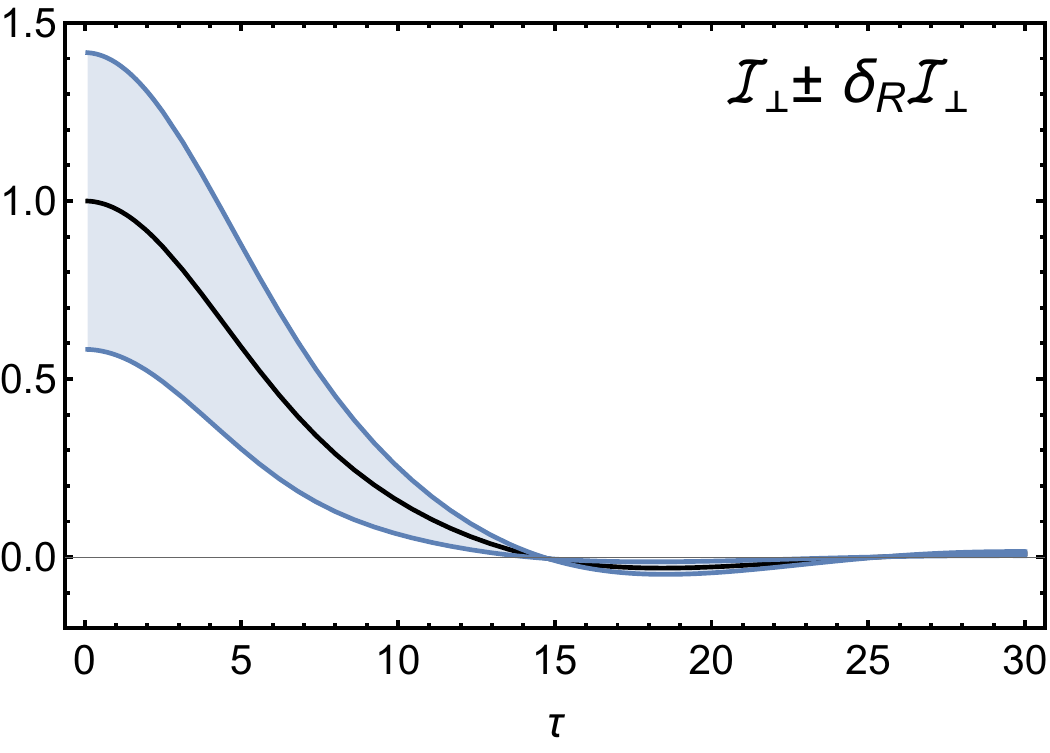}
 \includegraphics[width=0.32\textwidth]{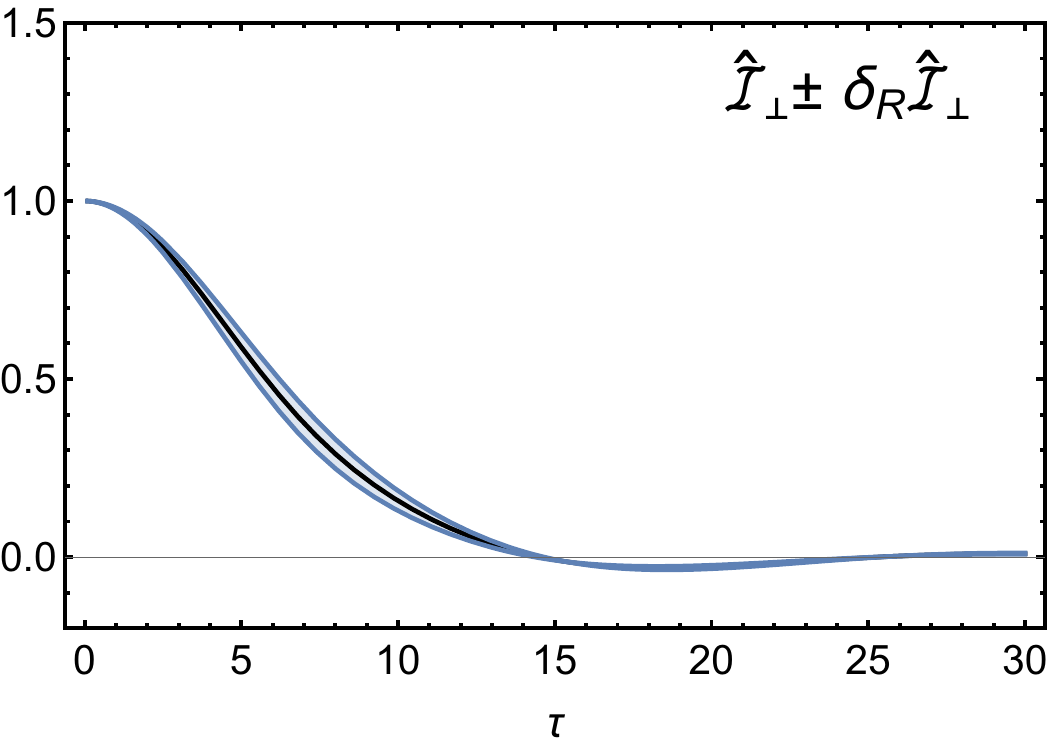}
 \includegraphics[width=0.32\textwidth]{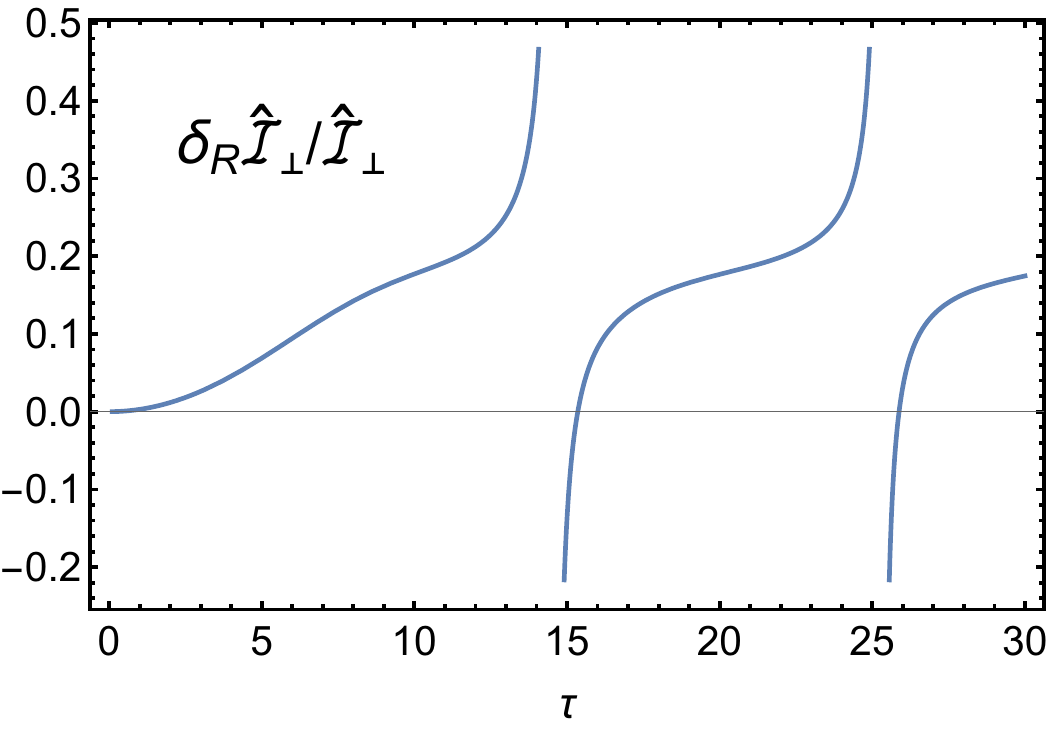}
\caption{Renormalon ambiguity for the position space (Ioffe-time)  valence quark distribution
for the model \eqref{GPDmodel} and the choice \eqref{norm-choice1} for the overall normalization,
corresponding to the quark-antiquark separation of the order of 1 fm.}  
\label{fig:ioffeplots}
\end{figure*}

The derivative $H'(x,\xi)$ 
is continuous at $x=\xi$ in the chosen model, see Fig.~\ref{fig:GPDmodel},
and has the following behavior:
\begin{align}
  H'(x,\xi) = \left\{
\begin{matrix}
 x\to \xiplus: & & 
H'(\xi,\xi) + a(x -\xi)^{1/2} 
+ \ldots
\\[1mm]
x\to \ximinus: & & 
H'(\xi,\xi) + \tilde a  (\xi-x)^{1} 
+ \ldots 
\end{matrix}  
\right.
\end{align}
where $a,\tilde a \sim 1/\xi^2$ in the small-$\xi$ limit. 
Only the second derivative is discontinuous. 
Thus for our model the delta-function terms
in Eq.~\eqref{qGPD-renormalons} for qGPDs do not contribute.
The continuity of the first derivative 
is reflected in the faster falloff of the corresponding 
position space GPD  $I(\tau,\xi) \sim 1/\tau^{5/2}$  instead of 
a generic behavior $I(\tau,\xi) \sim 1/\tau^{2}$ at $\tau\to\infty$, see the next section. 

This feature should be viewed as a certain shortcoming of the model:
We remind that discontinuity of the first derivative is an endemic feature of GPDs, it is 
generated invariably by the evolution even if the initial condition is continuous. 

In the following subsections we show the renormalon ambiguity that is representative for
the expected size of the nonperturbative power suppressed correction 
(for the chosen GPD model) for the position space, pGPD and qGPD, respectively.   
We plot the results for the transverse distributions, cf. Eq.~\eqref{long-trans}, which are 
the relevant ones for lattice calculations. The renormalon ambiguity for longitudinal distributions
is in all cases qualitatively similar, but is somewhat smaller in magnitude.

\subsection{Position space}

The renormalon ambiguity \eqref{Ihat} for the un-normalized $I(\tau,\xi)$ and the normalized $\widehat I(\tau,\xi)$
position space (Ioffe-time) distributions
for $\xi=0.3$ using the GPD model \eqref{GPDmodel} and the overall normalization 
\begin{align}
\mathcal{N}\,\Big(\Lambda^2z^2|v^2|\Big) &\mapsto 1
\label{norm-choice1}
\end{align}
corresponding to a large quark-antiquark separation of the order of 1 fm,
is shown in  Fig.~\ref{fig:ioffeplots} on the left two panels.
The shaded area corresponds to the interval spanned by $I(\tau,\xi) \pm \delta_RI(\tau,\xi)$
and similar for the normalized distributions.
In addition, the ratio $\delta_R\widehat I(\tau,\xi)/\widehat I(\tau,\xi)$ is plotted on the rightmost panel. 

One sees that the normalization to the correlation function at zero momentum leads to a strong 
reduction of the ambiguity and the expected higher-twist correction. The reason is that power
corrections depend only weakly of the hadron momentum as can be expected on general grounds. 
The renormalon ambiguity in the bubble chain approximation 
is essentially a measure of the phase space available for soft gluon emission in one-loop diagrams,
which can be made explicit using the gluon mass technique~\cite{Beneke:1994qe,Ball:1995ni},
cf. Eq.~\eqref{gluonmass}. Our calculation demonstrates that the phase space volume for soft emission 
does not decrease with the target momentum.

Note, however, that the ratio 
$\delta_R\widehat I(\tau,\xi)/\widehat I(\tau,\xi)$ diverges in the regions 
where the GPD \eqref{GPDmodel} crosses zero and changes sign. 
The reason is simply that the zeroes of $\delta_R\widehat I(\tau,\xi)$ 
and of $\widehat I(\tau,\xi)$ are shifted slightly with respect to one another.
It is not known to which extent this pattern of sign changes is specific to the 
chosen simple model, or more general. We conclude that position-space distributions at large
$\tau \gtrsim 10$ (in general $\tau \gg 1/\xi$) can be affected by large higher-twist effects and their interpretation 
within the leading twist factorization approximation should be considered with caution. 
 
\begin{figure*}[htb!]
\centering
 \includegraphics[width=0.32\textwidth]{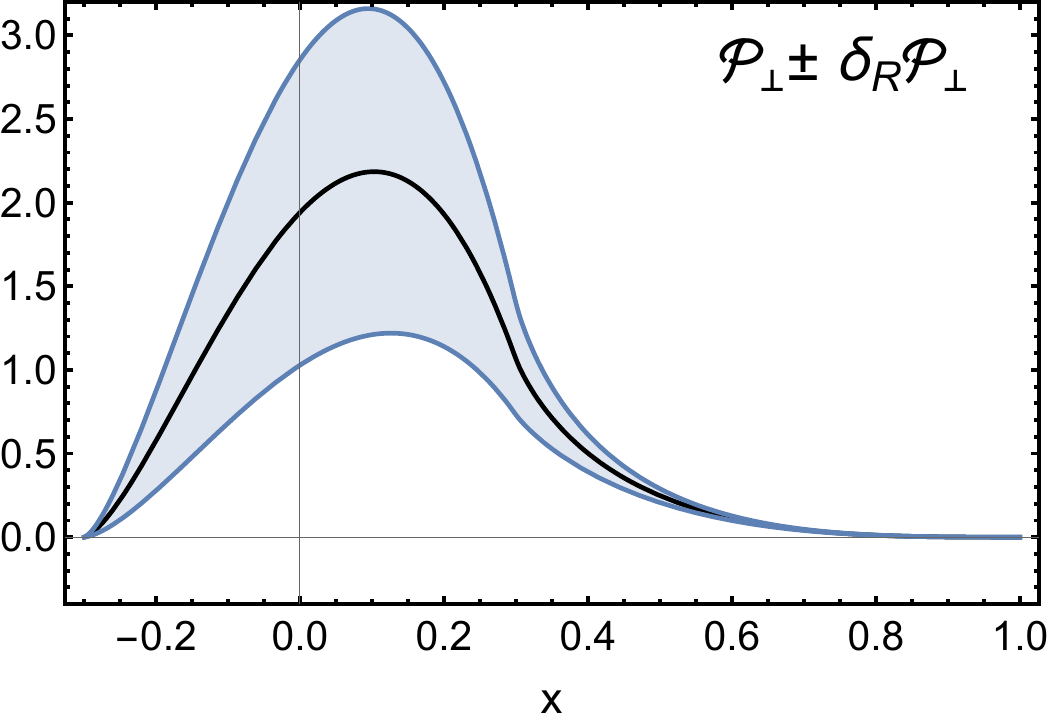}
 \includegraphics[width=0.32\textwidth]{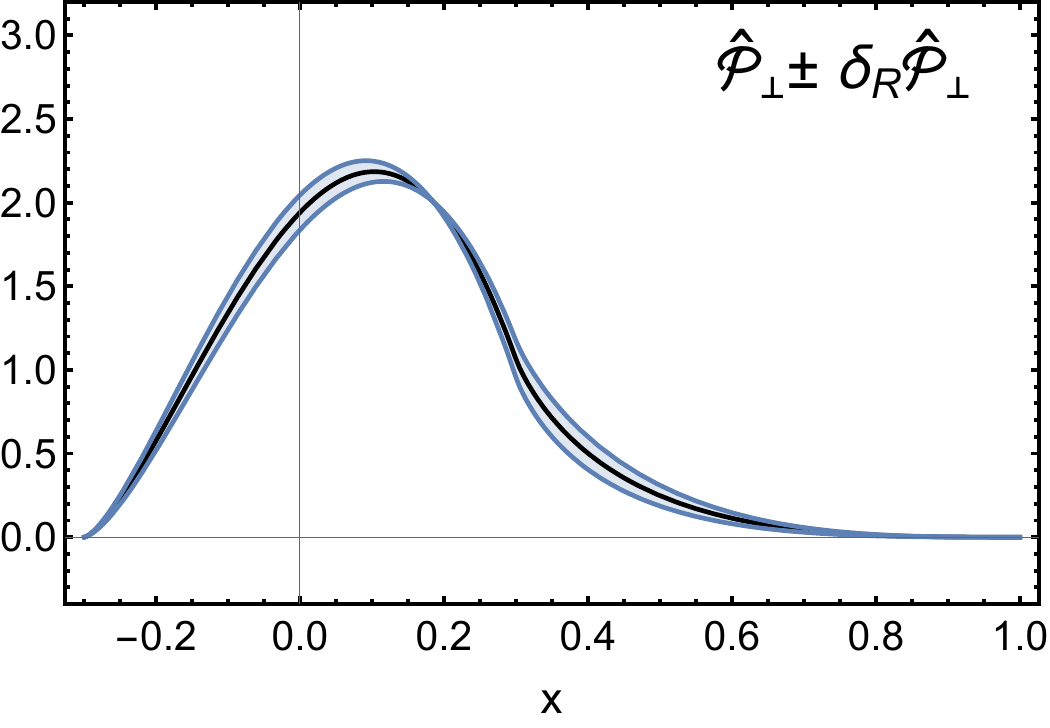}
 \includegraphics[width=0.32\textwidth]{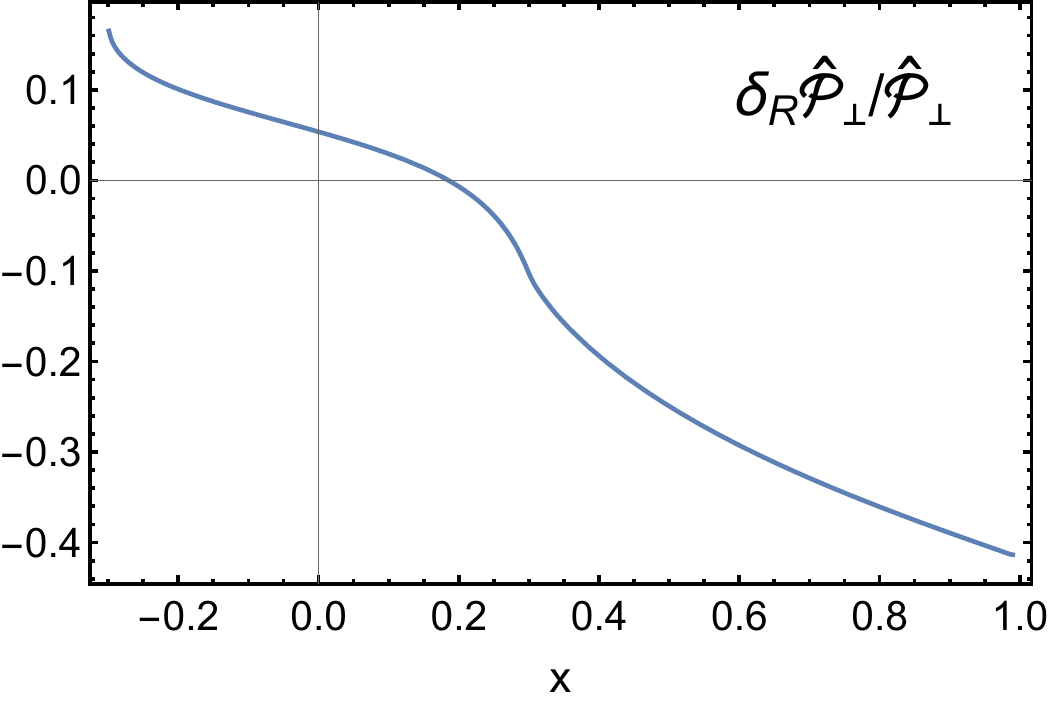}
\caption{Renormalon ambiguity for the  valence quark pseudo-GPD
for the model \eqref{GPDmodel} and the choice \eqref{norm-choice1} for the overall normalization,
corresponding to the quark-antiquark separation of the order of 1 fm.}  
\label{fig:pGPDplots}
\end{figure*}

\subsection{Pseudo-GPDs}
Our estimates for the renormalon ambiguity for valence quark pGPDs are plotted in Fig.~\ref{fig:pGPDplots}
for $\mathcal{P}_\perp(x,\xi)$, $\widehat{\mathcal{P}}_\perp(x,\xi)$ and  
for the ratio $\delta_R\widehat{\mathcal{P}}(x,\xi)/\widehat{\mathcal{P}}(x,\xi)$, 
for $\xi=0.3$ as a function of $x$,
from left to right, respectively.
The similar ratio for renormalized pGPDs $\delta_R{\mathcal{P}}(x,\xi)/{\mathcal{P}}(x,\xi)$
is shifted upwards by a constant $5/12$, cf. \eqref{hat-pGPD}, so that we do not plot it 
separately.

Similar to position space results, we observe a very strong reduction of the renormalon ambiguity 
for the pGPD normalized to zero momentum, cf. the left and the middle panel.
Remarkably, the residual ambiguity for the normalized pGPD is small for $x\simeq \xi$,
which is the most interesting region for DVCS phenomenology. 
The ratio $\delta_R\widehat{\mathcal{P}}(x,\xi)/\widehat{\mathcal{P}}(x,\xi)$ stays constant
at the end-point regions $x\to 1$ and $x\to -\xi$. This behavior is consistent with the study of 
renormalons in pPDFs in Ref.~\cite{Braun:2018brg}.

\subsection{Quasi-GPDs}
\label{sec:quasi-GPDs}

\begin{figure*}[htb!]
\centering
 \includegraphics[width=0.32\textwidth]{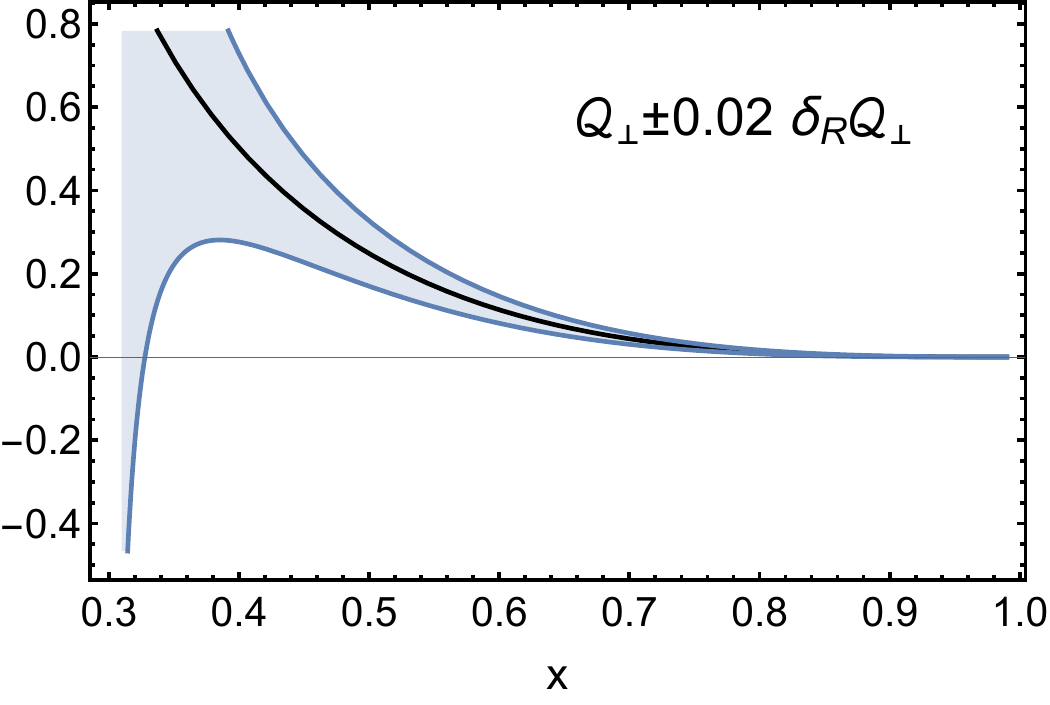}
 \includegraphics[width=0.32\textwidth]{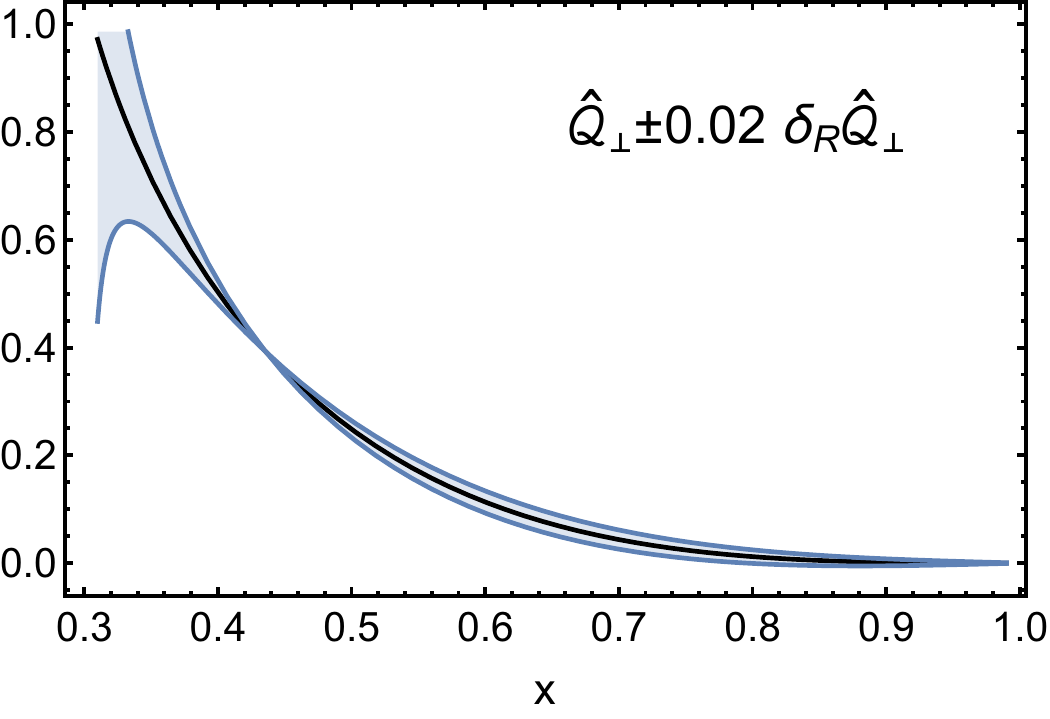}\\
 \includegraphics[width=0.32\textwidth]{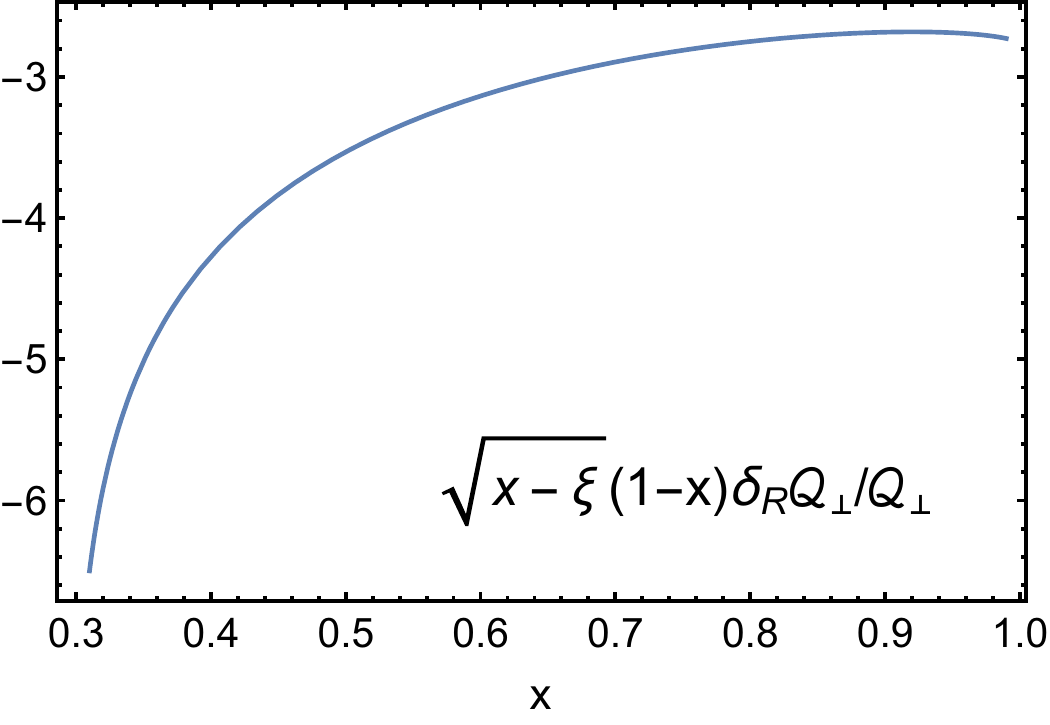}
\includegraphics[width=0.32\textwidth]{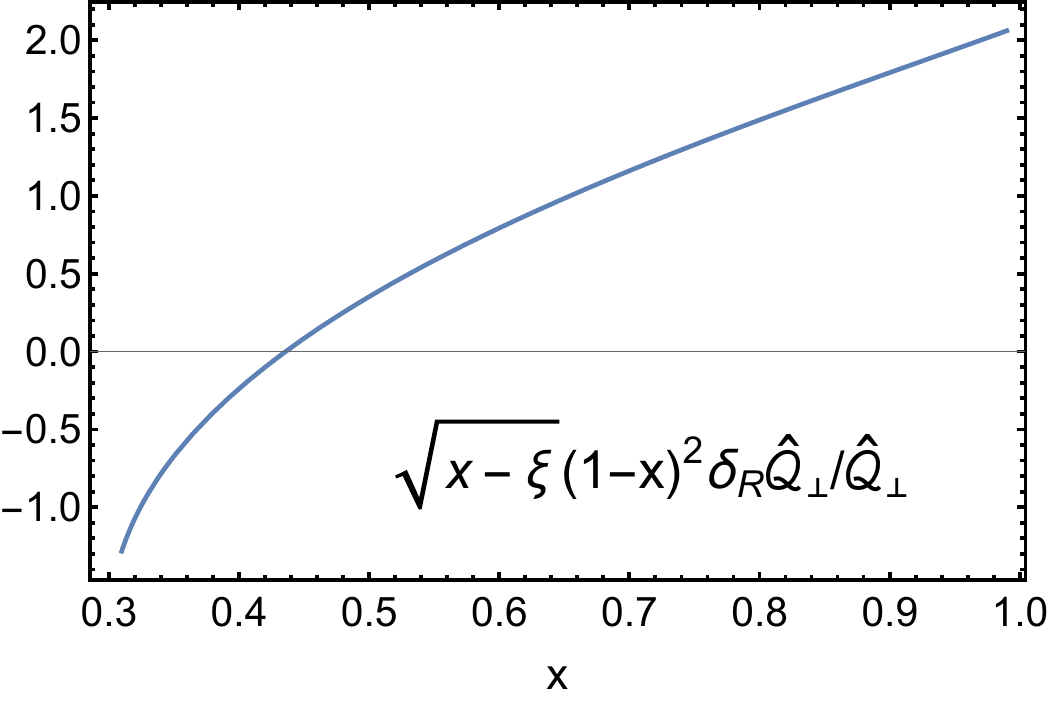}
\caption{Renormalon ambiguity for the  valence quark quasi-GPD in the DGLAP region $x>\xi$
for the model \eqref{GPDmodel} and the choice \eqref{norm-choice2} for the overall normalization,
corresponding to the target momentum approximately 1.5 GeV.}  
\label{fig:qGPD-DGLAP}
\end{figure*}

\begin{figure*}[htb!]
\centering
 \includegraphics[width=0.32\textwidth]{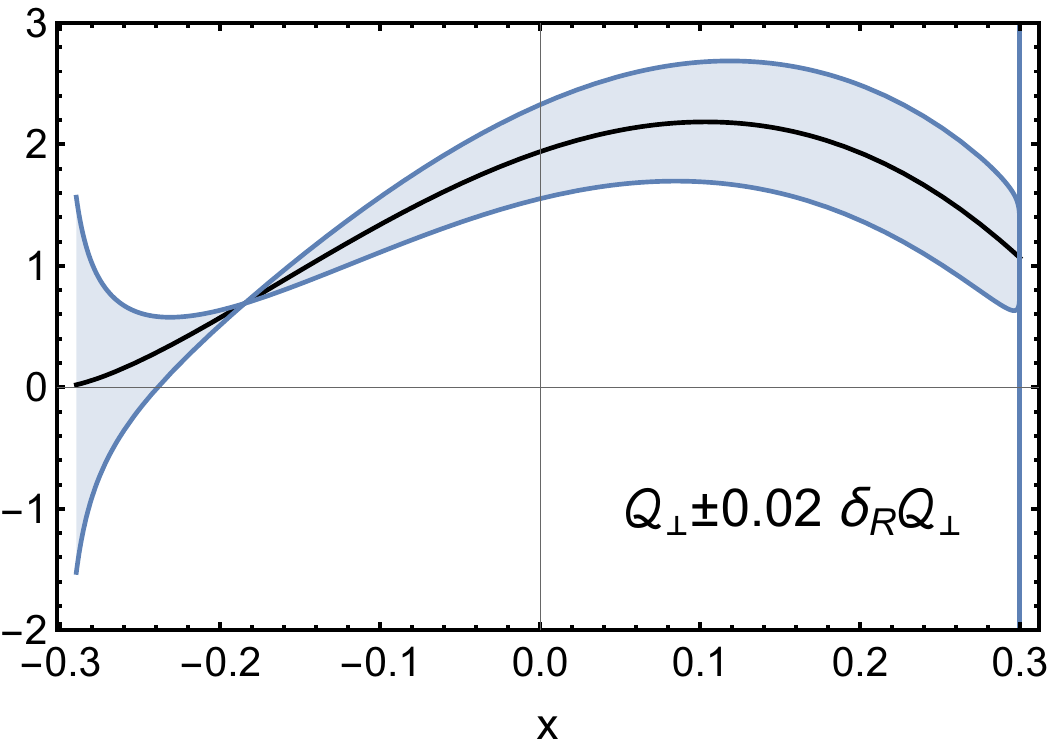}
 \includegraphics[width=0.32\textwidth]{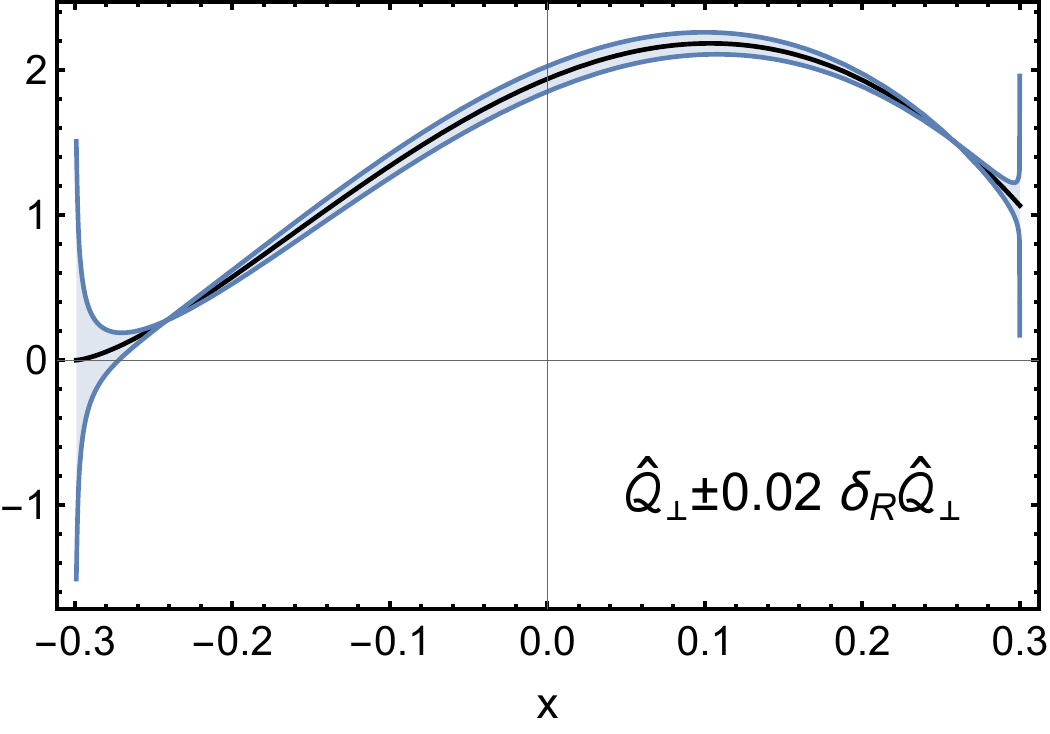}\\
 \includegraphics[width=0.32\textwidth]{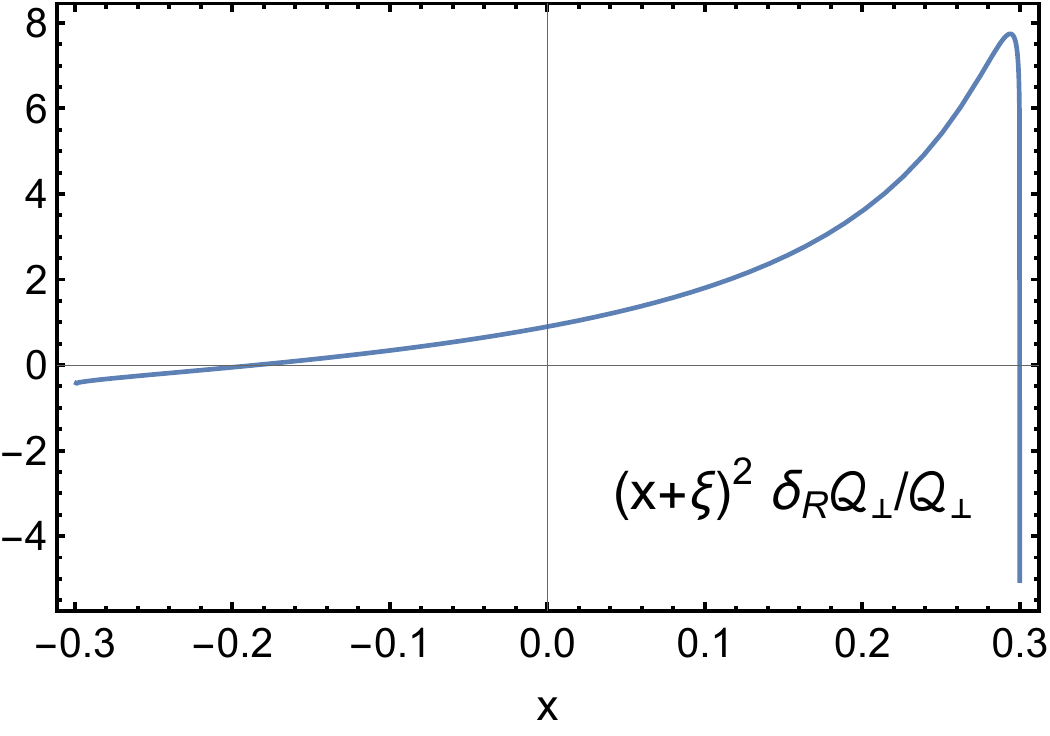}
\includegraphics[width=0.32\textwidth]{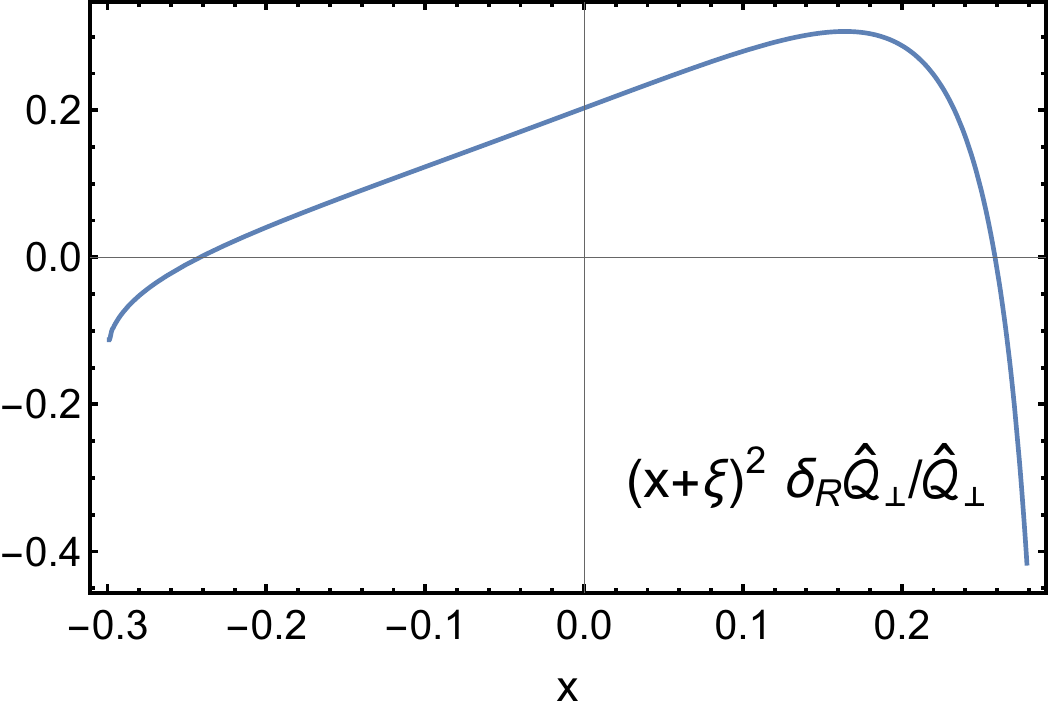}
\caption{Renormalon ambiguity for the valence quark quasi-GPD in the ERBL region $-\xi< x<\xi$
for the model \eqref{GPDmodel} and the choice \eqref{norm-choice2} for the overall normalization,
corresponding to the target momentum approximately 1.5 GeV.}  
\label{fig:qGPD-ERBL}
\end{figure*}

Our results for qGPDs \eqref{qGPD-ambiguity} are shown in Fig.~\ref{fig:qGPD-DGLAP}
and Fig.~\ref{fig:qGPD-ERBL} for the DGLAP ($x>\xi$) and ERBL ($-\xi<x<\xi$)
regions, respectively.  
For this calculation we set the overall normalization factor to
\begin{align}
\mathcal{N}\,\left(\frac{\Lambda^2|v^2|}{(vP)^2}\right) \mapsto 0.02\,. 
\label{norm-choice2}
\end{align} 
This choice, assuming $\Lambda \simeq 200\,\text{MeV}$, corresponds to the 
target momentum of roughly $(vP) \simeq 1.5\,\text{GeV}$, which is a typical value 
accessible in present day lattice calculations.

We find the following behavior of the renormalon ambiguity for the ratios 
$\delta_R\mathcal{Q}(x,\xi)/\mathcal{Q}(x,\xi)$ in the 
end-point regions:
\begin{itemize}
  \item  $1/(1-x)$ enhancement for $x\to 1$
  \item  $1/\sqrt{x-\xi}$  enhancement for $x\to \xi$ from above
  \item  $\ln(\xi-x)$  enhancement for $x\to \xi$ from below
  \item  $1/(x+\xi)^2$ enhancement for $x\to -\xi$ from above
\end{itemize}
The $\sim 1/\sqrt{x-\xi}$  behavior of the renormalon ambiguity for $x\to \xi_>$ 
is specific for our model which corresponds to the valence quark PDF $q(x)\sim 1/\sqrt{x}$. A generic Regge-type 
small-$x$ behavior $q(x)\sim 1/x^p$ leads to the $1/(x-\xi)^p$ enhancement.  

Similar to pGPDs, we find that the normalization to the position-space correlation function at zero 
momentum leads to the strong reduction of the renormalon ambiguity, 
$\delta_R\widehat{\mathcal{Q}}(x,\xi)/\widehat{\mathcal{Q}}(x,\xi) \ll \delta_R\mathcal{Q}(x,\xi)/\mathcal{Q}(x,\xi)$,
apart from the large-$x$ region 
where the ambiguity is instead parametrically enhanced, $1/(1-x)\mapsto 1/(1-x)^2$. 
This enhancement is the same as found for qPDFs/pPDFs in Ref.~\cite{Braun:2018brg}, which is expected
as GPDs and PDFs coincide for $x\gg \xi$. 

%
\section{Power corrections at  $x\to\xi$}
\label{sec:x=xi}
%
The kinematic points $x=\pm \xi$ correspond to vanishing longitudinal momentum 
of the quark or the antiquark in the target and can only be included in 
collinear factorization by a contour deformation as 
explained in \cite{Radyushkin:1997ki,Collins:1998be} for DVCS.
It is therefore not surprising that this point 
requires careful treatment.

The appearance of the delta-function contributions to the qGPDs
in Eq.~\eqref{qGPD-renormalons} can be explained as follows. 
For simplicity consider valence quark GPD which vanishes
for $x<-\xi$, as in the above model.
From the definition in \eqref{GPDdef}
\begin{align}
I(\tau,\xi) &= -\frac{1}{\tau^2} \int_{-\xi}^1\!dx\,  H(x,\xi) \frac{d^2}{dx^2} e^{i\tau x}
\notag\\&=
  \frac{1}{\tau^2} \Big[H'(\ximinus,\xi)-H'(\xiplus,\xi)\Big] e^{i\tau\xi} 
\notag\\&\quad
-  \frac{1}{\tau^2} \left(\int_{-\xi}^{\ximinus}+  \int_{\xiplus}^1\right)dx\, e^{i\tau x} H''(x,\xi)
\end{align}
where we assumed that $H(x,\xi)$ is continuous and vanishes sufficiently fast at the end points,
but has a cusp at $x=\xi$. The remaining integral is finite and decreases at $\tau\to\infty$ 
provided $H''(x,\tau)$ does not have a strong singularity $1/(x-\xi)^p$ with $p\ge 1$, which
is the case for all existing models based on the double-distribution representation.
Therefore, in the general case, for large distances
\begin{align}
I(\tau,\xi) &\simeq \frac{1}{\tau^2} e^{i\tau\xi} [H'(\xi\!+\!0,\xi)- H'(\xi\!-\!0,\xi)]\,. 
\label{large-tau}
\end{align}
(Extra boundary terms contributions may need to be added 
if $H'(x,\xi)$ does not vanish at $x\to 1$ and $x\to -\xi$. )

The Fourier transform of \eqref{large-tau} in $\tau$ for fixed $z$ and hence the corresponding contribution 
to the pGPDs are well defined. For qGPDs, however, the $1/\tau^2 \sim  1/z^2$ suppression (for fixed $(vP)$) 
is lifted by the $z^2$ factor in the renormalon contribution in Eq.~\eqref{Ihat}. Hence the Fourier integral in $z$
is dominated by large distances, giving rise to the delta-function terms in~\eqref{qGPD-renormalons}.  
It is easy to convince oneself that further renormalons  $\sim z^4, z^6,\ldots$ 
lead to increasingly more singular contributions that have to be resummed. This situation is reminiscent of 
exclusive reactions in which case contributions of higher-twist LCDAs have to be supplemented by taking 
into account the so-called end-point contributions coming from large distances.

In other words, the limits $(vP)\to \infty$ and $x\to \xi$ do not commute. 
Our previous calculation assumed a short-distance expansion of the renormalon ambiguity in powers of  $\sim z^2, z^4,\ldots$
alias $1/(vP)^2, 1/(vP)^4\ldots$ for {\it fixed} $x-\xi$, whereas in this section we take the limit 
$x\to \xi$ first, and only then expand in powers of the target momentum. 
To this end we study the renormalons in qGPDs directly, 
avoiding the expansion at small $z^2$ at the intermediate step.
It is easy to see that the terms of interest originate from the vertex corrections Fig~\ref{fig:diagrams}a,b
only (the terms involving the hypergeometric function in Eq.~\eqref{start}).  
Omitting the remaining terms and using the definition of a qGPD in \eqref{def:qpGPD} we obtain
\begin{widetext}
\begin{align}
B[\mathcal{Q}(x,\xi,(vP))](w) & =
2 C_F e^{5/3w} \left(\frac{\mu\,|v|}{2 (vP)}\right)^{2w}
\frac{\Gamma(-w)}{\Gamma(w+2)}   
\int_{-\infty}^\infty\!\!\frac{d\tau}{2\pi}\,e^{-ix\tau} (\tau^2)^w 
\int_0^1 d\alpha\,\alpha^{1+w}  {}_2F_1(1,2-w,2+w,\alpha)
\notag\\&\quad
\times
\int_{-1}^1\!dy\, H(y,\xi) 
\biggl[ e^{i\tau(\alpha y+\bar\alpha\xi)} + e^{i\tau(\alpha y-\bar\alpha\xi)} - 2 e^{i\tau y}\biggr]+\ldots
\label{start2}
\end{align}
The Fourier integral can be taken: 
\begin{align}
 \Gamma(-w) \int_{-\infty}^\infty\frac{d\tau }{2\pi}\, e^{iq\tau} [\tau^2]^{w} 
& = \frac{ \Gamma(w+1/2)}{[q^2]^{w+1/2}} \frac{1}{\sqrt{\pi}} 4^w,
\end{align}
so that one gets
\begin{align}
B[\mathcal{Q}(x,\xi,(vP))](w) & =
\frac{2 C_F}{\sqrt{\pi}} e^{5/3w} \left(\frac{\mu\,|v|}{(vP)}\right)^{2w}
\frac{\Gamma(w+1/2)}{\Gamma(w+2)}   
\int_0^1 d\alpha\,\alpha^{1+w}  {}_2F_1(1,2-w,2+w,\alpha)
\notag\\&\quad
\times
\int_{-1}^1\!dy\, H(y,\xi) 
\biggl[ \left(
\frac{1}{[(x-\alpha y -\bar\alpha \xi)^2]^{w+1/2}}
- \frac{1}{[(x-y)^2]^{w+1/2}}\right) + (\xi\leftrightarrow -\xi) \biggr]+\ldots
\label{start3}
\end{align}
One can show that for $x\slashed{=}\pm \xi$ the leading renormalon singularity in this expression is located at $w=1$
and reproduces the expression in Eq.~\eqref{qGPD-renormalons}. For $x=\pm\xi$ the result is different.
At this special point the remaining integrals factorize.
Using
\begin{align}
  \int_0^1 d\alpha\,\alpha^{1+w}  {}_2F_1(1,2-w,2+w,\alpha) 
\left(\frac{1}{\alpha^{2w+1}} -1\right)  = \frac{1}{1-w}
\end{align}  
we obtain
\begin{align}
B[\mathcal{Q}(\xi,\xi,(vP))](w) & =
\frac{2 C_F}{\sqrt{\pi}} e^{5/3w} \left(\frac{\mu\,|v|}{(vP)}\right)^{2w}
\frac{\Gamma(w+1/2)}{\Gamma(w+2)}   
\frac{1}{1-w}
\int_{-1}^1\!dy\, H(y,\xi) 
\biggl[\frac{1}{[(y-\xi)^2]^{w+1/2}} +  \frac{1}{[(y+\xi)^2]^{w+1/2}}\biggr]+\ldots
\label{start4}
\end{align}
Singularities in the Borel plane, apart from the prefactor, can arise from vanishing of the 
denominator in the remaining integral. We write
\begin{align}
 H(y,\xi) &= H(\xi,\xi) + (y-\xi)H'(\xi_{>(<)},\xi) + \delta H(y,\xi)
\end{align}
where $\delta H = o(y-\xi)$ (i.e. vanishes faster than $y-\xi$). Then
\begin{align}
 \int_{-1}^1\!dy\,\frac{H(y,\xi)}{[(y-\xi)^2]^{w+1/2}}
&=
- \frac{1}{2w} H(\xi,\xi) \Big[(1-\xi)^{-2w} +(1+\xi)^{-2w}\Big]
+ \frac{1}{1-2w} \Big[H'(\xiplus,\xi)(1-\xi)^{1-2w} - H'(\ximinus,\xi)(1+\xi)^{1-2w}\Big]  
\notag\\&\quad
+ \int_{-1}^1\!dy\,\frac{\delta H(y,\xi)}{[(y-\xi)^2]^{w+1/2}}.
\label{55}
\end{align}
\end{widetext}
The first term $\sim 1/w$ in this expression is removed by the overall renormalization,
and the leading renormalon ambiguity stems from a new singularity at $w=1/2$ corresponding to a
power correction $1/(vP)^1$, i.e. suppressed by only one power of the large momentum.
Note that this contribution is proportional to the discontinuity in the first derivative, which,
as already mentioned above, is an endemic feature of quark GPDs.
 
The remainder $\delta H(y,\xi)$ is generally $\sim (\xi-y)^2$ in the ERBL region $y<\xi$, 
whereas its behavior in the DGLAP region appears to be related to the small-$x$ behavior 
of the corresponding PDF. For the model \eqref{GPDmodel}  $\delta H(y,\xi) \stackrel{y\to \xiplus}{\sim} (y-\xi)^{\lambda}$
corresponding to $q(x) \sim x^{\lambda-2}$.
In this case
\begin{align}
    \int_{-1}^1\!dy\,\frac{\delta H(y,\xi)}{[(y-\xi)^2]^{w+1/2}} \sim \frac{c_1}{\lambda/2-w} + \frac{c_2}{1-w} 
\end{align}
where the first term with $c_1 \sim 1/\xi^2$  originates from the DGLAP region only. Taking into account an overall $1/(1-w)$ prefactor
in Eq.~\eqref{start4},  the second contribution gives rise to a double pole $1/(1-w)^2$ corresponding to a 
power correction $\sim \ln(vP)/(vP)^2$, whereas the first contribution indicates existence of a fractional power correction $\sim 1/(vP)^{\lambda}$  
with, for realistic models, $1 < \lambda <2$. The possibility of fractional power corrections was previously discussed 
by Manohar and Wise~\cite{Manohar:1994kq} in connection with event shape variables in $e^+e^-$ annihilation.    

To summarize, qGPDs at the kinematic point $x=\xi$ (and $x=-\xi$ for non-valence GPDs) 
in the single bubble chain approximation have the following singularities in the Borel plane:
\begin{enumerate}
\item A pole at $w=1/2$ corresponding to a power correction $1/(vP)^1$;
\item A pole at  $w=\lambda/2$, $1<\lambda < 2$ corresponding to a power correction $1/(vP)^\lambda$;
\item A double pole at  $w=1$, corresponding to a power correction $1/(vP)^2$, enhanced by an additional logarithm.
\end{enumerate} 
Note also that $B[\mathcal{Q}(\xi,\xi,(vP))](w)$ contains an UV renormalon singularity at $w=-1/2$ corresponding to a Borel-summable 
(sign-alternating) perturbative series. This singularity does not imply any additional nonperturbative corrections, 
but signals existence of large perturbative contributions to the qGPD coefficient function at $x=\xi$ at high orders.

The appearance of new singularities in the Borel plane for qGPDs at $x\to\xi$ 
can also be understood in a different language. 
Note that the $1/(Pv)^2$ power correction (= renormalon ambiguity) to the qGPD in 
Eq.~\eqref{qGPD-renormalons}  is a distribution in mathematical sense, and it also contains
a divergent at $x\to \xi$ contribution $\sim 1/\sqrt{x-\xi}$ from the DGLAP region, 
cf. Sec.~\ref{sec:quasi-GPDs} and Fig.~\ref{fig:qGPD-DGLAP}.
In order to understand the impact of these corrections one has to consider their action on suitable 
test functions. To this end, consider a qPDF smeared over a narrow interval $x-\xi \sim \Lambda/P$,
\begin{align}
\widetilde{\mathcal{Q}}(x,\xi) = \int dx'\, \Theta(x'-x)\mathcal{Q}(x',\xi)\,, &&
\int dx\,\Theta(x) =1\,. 
\end{align}
For definiteness, one can use Gaussian smearing
\begin{align}
 \Theta(x) = \frac{1}{\sqrt{\pi}}\frac{(Pv)}{\Lambda} \exp\Big(- (Pv)^2 x^2/\Lambda^2 \Big).
\end{align}
It is easy to see that if this smearing is applied to Eq.~\eqref{qGPD-ambiguity}, 
the term with a $\delta$-function in Eq.~\eqref{qGPD-renormalons} 
gets promoted to a $1/(vP)$ correction with the 
coefficient consistent with Eq.~\eqref{55}, and the term $\sim 1/\sqrt{x-\xi}$ from the 
asymptotic expansion of the integral at $x\to \xi$ (for our GPD model) 
produces a $1/(Pv)^{3/2}$ contribution, in exact correspondence to the findings in this
section.\\   
The lesson to be learned from this comparison is that qPDF approach remains viable in
the $x\to\xi$ region despite the apparently divergent contributions in this limit
in the $1/(Pv)^2$ expansion. The nonperturbative corrections remain 
finite with only the power suppression changing from $1/(Pv)^2$ at 
$x-\xi = \mathcal{O}(1)$ to  $1/(Pv)$ at $x-\xi \lesssim \mathcal{O}(\Lambda/(Pv))$.
This situation is much better compared to what happens in qPDFs at small Bjorken x,
in which case the nonperturbative corrections explode and the qPDF approach is not applicable.\\
A final remark is that a similar ``transmutation'' of the power of the power correction 
$1/Q^2\to 1/Q$ was found previously in single inclusive particle production in $e^+e^-$ annihilation for the 
longitudinal and transverse cross sections integrated over the small momenta region, see
Refs.~\cite{Beneke:1996rz,Dasgupta:1996ki,Beneke:1997sr}.

%
\section{Conclusions}
\label{sec:summary}
%
We have calculated the renormalon ambiguity in qGPDs and pGPDs in the single bubble-chain approximation, 
which is a generalization and extension of the previous results on qPDFs/pPDFs from Ref.~\cite{Braun:2018brg}
and pion pseudo-LCDAs from Ref.~\cite{Braun:2004bu} to a more general kinematics. Our results are collected
in analytic form in Sec.~\ref{sec:analytics}, complemented by  a numerical study for a simple GPD model in
Sec.~\ref{sec:numerics}.    
We find that the normalization to the position-space correlation function at zero 
momentum leads to a strong reduction of the renormalon ambiguity for both pPDFs and qGPDs, apart from the large-$x$ region 
where the ambiguity is instead parametrically enhanced, $(1-x)^1 \mapsto (1-x)^0$ and $1/(1-x)\mapsto 1/(1-x)^2$ for pGPDs and qGPDs,
respectively. The same  enhancement was found in Ref.~\cite{Braun:2018brg} for qPDFs/pPDFs.

The most interesting difference between pGPDs and qGPDs appears to be the behavior of the renormalon ambiguity 
in the kinematic region $x\to \xi$ which corresponds to zero longitudinal momentum carried by one of the partons and 
is most relevant for DVCS phenomenology. The renormalon ambiguity for pGPDs in the $x\simeq \xi$ region proves to be 
small, see Fig.~\ref{fig:pGPDplots}, whereas for qGPD it is large and depends on the direction in which the limit
$x\to \xi$ (from above or from below) is taken. Moreover, for qGPDs, the limits $x\to \xi$ and the target momentum $(vP)\to \infty$ do not commute.
Expanding first in $1/(vP)$ produces a series of renormalons involving the $\delta$-function $\delta(x-\xi)$ and its derivatives 
\eqref{qGPD-renormalons}, 
whereas setting $x\to\xi$ and then expanding in powers of  $1/(vP)$ reveals instead two new IR renormalon singularities:
The first one (leading) corresponds to a power correction suppressed by the first power $1/(vP)^1$, and the second one  
indicates the existence of a correction with noninteger power suppression $1/(vP)^{2-p}$ for the assumed small-$x$ quark PDF behavior
$q(x) \sim x^{-p}$, with $0<p<1$. Note that the $1/(vP)^1$ power correction in qGPDs at $x=\xi$ is unrelated to  
the UV renormalon ambiguity in the off-light cone Wilson line which is assumed to be eliminated by a proper renormalization/normalization
procedure. Instead, it is generated by the discontinuity if the $x$-derivative of the quark GPD at $x=\xi$ that is necessitated by 
the evolution equation.   

\section*{Acknowledgments}

This work was supported by the Research Unit FOR2926 funded by the Deutsche
Forschungsgemeinschaft (DFG, German Research Foundation) under grant 409651613.
J.S. was also supported in part  by the U.S. Department of Energy
through Contract No. DE-SC0012704 and by Laboratory
Directed Research and Development (LDRD) funds from
Brookhaven Science Associates.


%

%

\bibliographystyle{apsrev}
\bibliography{ref_twist4}%

\end{document}